\documentclass[draft]{agujournal2019}
\usepackage{url} %this package should fix any errors with URLs in refs.
\usepackage{lineno}
\usepackage[inline]{trackchanges} %for better track changes. finalnew option will compile document with changes incorporated
\usepackage{soul}
\usepackage{amsmath}
\usepackage{gensymb}
\usepackage{color}

\draftfalse

\journalname{Geophysical Research Letters}

\makeatletter
\def\ps@titlepage{%
    \def\@oddhead{}
    \def\@evenhead{}
    \def\@oddfoot{\reset@font\hfil\thepage\hfil}%
    \def\@evenfoot{\reset@font\hfil\thepage\hfil}%
}

\def\ps@standard{%
    \def\@oddhead{}
    \def\@evenhead{}
    \def\@oddfoot{\reset@font\hfil\thepage\hfil}
    \def\@evenfoot{\reset@font\hfil\thepage\hfil}
}

\makeatother
\begin{document}

\title{Exploring the conditions conducive to convection within the Greenland Ice Sheet}% | Local convection within the Greenland Ice Sheet may constrain ice rheology | Evidence for local convection within the Greenland Ice Sheet}

\authors{Robert Law \affil{1, 2}, Andreas Born \affil{1, 2}, Philipp Voigt \affil{1, 2}, Joseph A. MacGregor \affil{3}, Claire Marie Guimond \affil{4}}

\affiliation{1}{Department of Earth Sciences, University of Bergen, Norway}
\affiliation{2}{Bjerknes Centre for Climate Research, Bergen, Norway}
\affiliation{3}{Cryospheric Sciences Laboratory, NASA Goddard Space Flight Center, Greenbelt, USA}
\affiliation{4}{Atmospheric, Oceanic, and Planetary Physics, University of Oxford, UK}

\correspondingauthor{Robert Law}{robert.law@uib.no}

\begin{keypoints}
\item Convection may explain the distribution of large basal plumes observed within the Greenland Ice Sheet
\item Horizontal shearing and high snowfall inhibit convection, while stable, thick, low-viscosity ice promotes convection
\item The presence or absence of these plumes may provide crucial constraints on ice-sheet rheology
\end{keypoints}

\begin{abstract}

Large plume-like features within the Greenland Ice Sheet disrupt radiostratigraphy and complicate the use of isochrones in reconstructions of past ice dynamics. Here we use numerical modeling to test the hypothesis that convection is a viable mechanism for the formation of the large (\(>\)1/3 ice thickness) englacial plume-like features observed in north Greenland. Greater horizontal shear and snow accumulation impede formation of convection plumes,  while stable and softer ice encourages them. These results potentially explain the dearth of basal plumes in the younger and higher-accumulation southern ice sheet. Leveraging this mechanism to place bounds on ice rheology suggests that -- for north Greenland -- ice viscosity may be \(\sim\)9-15 times lower than commonly assumed. Softer-than-assumed ice there implies significantly reduced basal sliding compared to standard models. Implementing a softer basal ice rheology in numerical models may help reduce uncertainty in projections of future ice-sheet mass balance. 

\end{abstract}

\section*{Plain Language Summary}

How ice deforms in response to a given stress state (its rheology) is a crucial yet poorly constrained relationship in the numerical models used to project future ice-sheet evolution. Here we suggest that large plumes of disturbed ice near the base of the ice sheet visible in radar images may form through thermal convection -- a process that can help us better understand ice rheology. The occurrence of convection depends strongly on ice thickness, ice temperature and softness, snowfall rates, and how fast the ice is flowing. These dependencies may explain why we see these plumes mostly in north Greenland, where the ice is old (therefore soft) and slow-moving but not in the south where faster ice flow and higher snowfall prevent convection. Our analysis suggests northern Greenland's ice may be 9--15 times softer than typically used in models. This means that internal ice deformation, rather than sliding at the base, may be the main way ice moves in these regions. If convection is the only way these plumes can form, using these updated ice properties in models will improve our ability to accurately project future changes in ice sheets.

\section{Introduction}

The Greenland Ice Sheet (GrIS) is losing mass at a rate that threatens coastal communities and infrastructure \cite{Otosaka2023Mass2020}, with numerical models predicting accelerating mass loss throughout the 21\textsuperscript{st} century and beyond \cite{Aschwanden2022CalibratedSheet}. Central to these predictions are descriptions of ice-sheet flow, which in turn require accurate parameterisations of ice rheology. However, despite its critical control on ice rheology, there are few reliable in-situ constraints for the value for the enhancement factor \(E\), which captures the rheological influence of variations in grain size, impurity content, and ice fabric. $E$ is defined as \(\dot{\epsilon}_m / \dot{\epsilon}_o\) \,, where \(\dot{\epsilon}_m\) is the measured strain rate and \(\dot{\epsilon}_o\) is the strain rate predicted by the standard Nye-Glen isotropic flow law (see Materials and Methods). $E$ varies based on deformation type but is often assumed to be $\sim$4--6 for the GrIS when combined with a flow exponent \(n = 3\) \cite{Cuffey2010TheGlaciers}. However, \(E\) is inferred to be up to 12 in Antarctic shear margins \cite{Echelmeyer1994TheStream} and even 120 in mountain glaciers \cite{Echelmeyer1987DirectTemperatures}. Deviations in \(E\) away from its appropriate value will result in compensatory errors in basal traction inversions, which ultimately negatively influence the accuracy of numerical models of ice sheets over time \cite{Berends2023CompensatingResponse}. %Any possibility to ascertain \(E\) \textit{in-situ} is therefore welcome. %Neither is there a particularly firm consensus for value the exponent, \(n\), in the Nye-Glen isotropic flow law should take, with recent suggestions that \(n=4\) is more appropriate in many ice-sheet settings than the default \(n=3\) \cite{Bons2018GreenlandMotion, Ranganathan2024ASheets}. %Given the natural variability in ice fabric and impurity content it is furthermore likely that both of these values exhibit at least some variability both in different glacial settings and within the same ice body.

\begin{figure}
\noindent\includegraphics[width=0.9\textwidth]{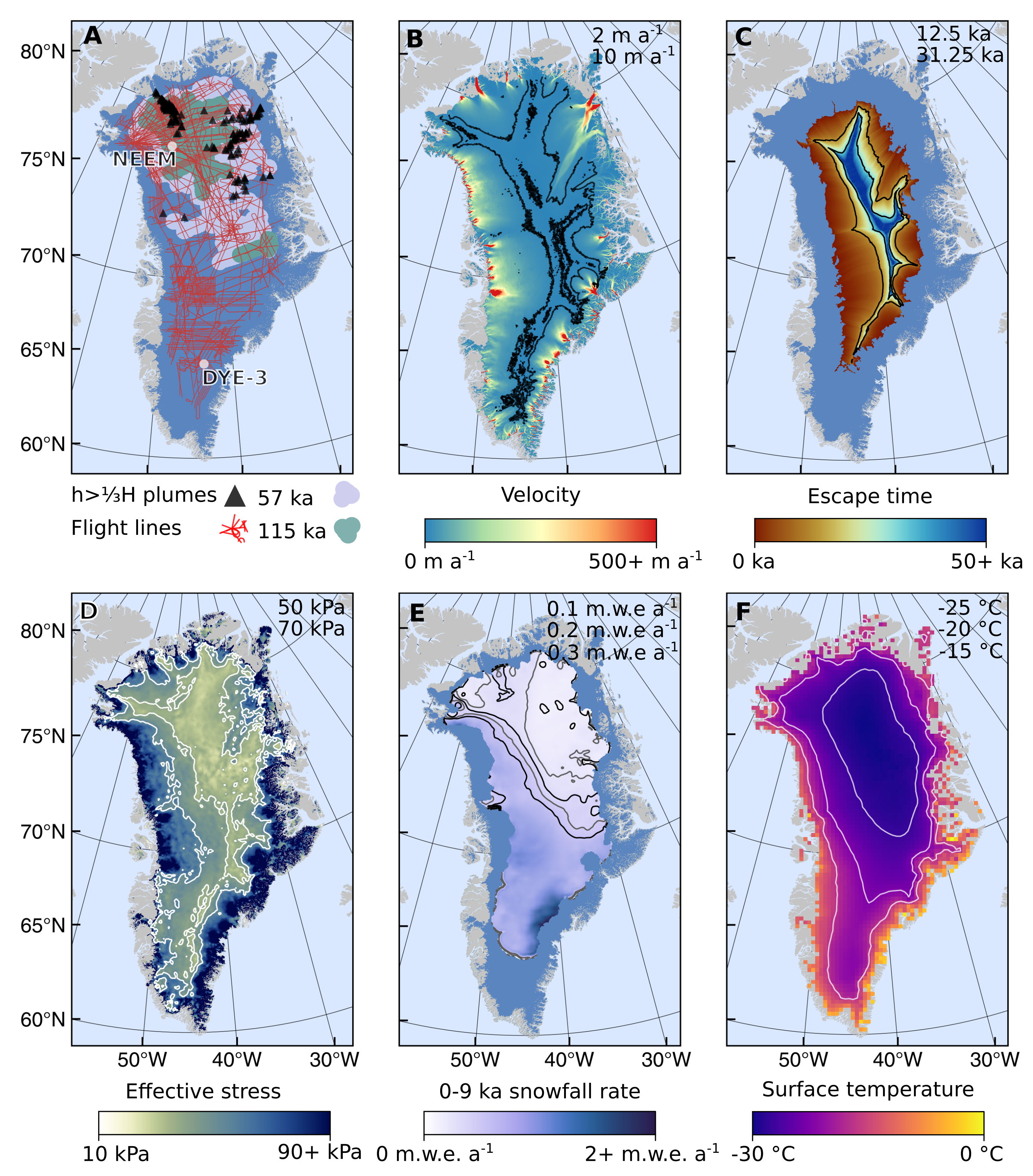}
\caption{Relevant GrIS properties when considering convection. Contour values are given in the top right of each panel. \textbf{A} Location of large plumes from \citeA{LeysingerVieli2018BasalStratigraphy}, NEEM and DYE-3 boreholes, flight lines, and detection of 57 and 115 ka age ice layers from \citeA{Macgregor2015RadarSheet}. \textbf{B} Escape time required to get reach the 2,000 m ice-thickness contour using NASA MEaSUREs ITS\_LIVE velocity data and a shape factor of 0.8 to better approximate column-averaged velocity (e.g. \citeA{Whillans1977TheAntarctica}). \textbf{C} Surface velocity using the same source as \textbf{B}. \textbf{D} Effective stress at 5/14 depth obtained from ISSM run. Fig. S2 shows a 3D view of this effective stress. \textbf{E} Averaged accumulation rate from \citeA{MacGregor2016HoloceneSheet} for 0-9 ka. \textbf{F} Mean annual temperature from RACMO averaged over 1959-2019 \cite{Noel2018Modelling19582016} Background data from QGreenland \cite{Moon2022QGreenland:GIS}}. 
\label{fig:location}
\end{figure}

\begin{figure}
\noindent\includegraphics[width=0.9\textwidth]{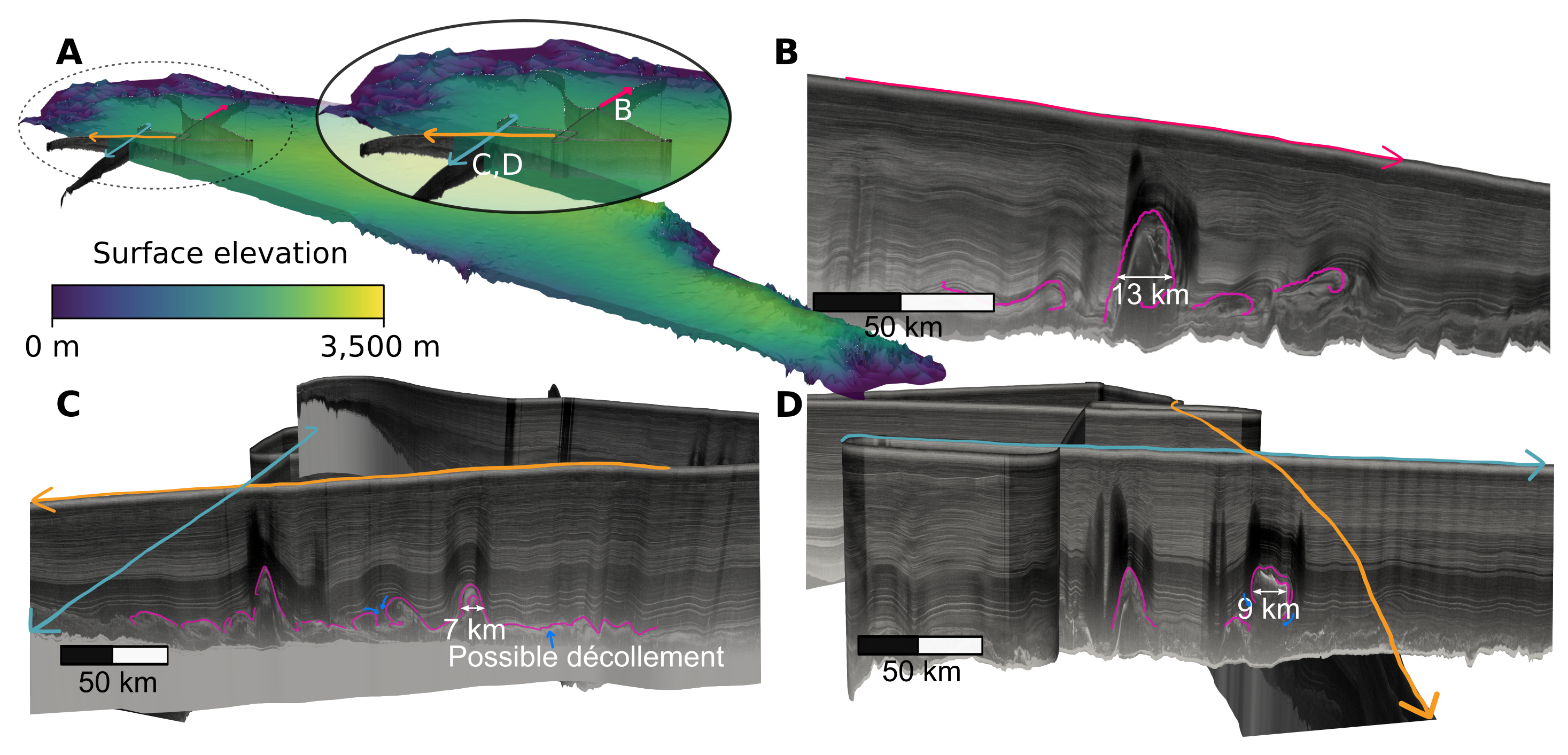}
\caption{Example plume features in north Greenland. \textbf{A} Oblique view of the GrIS with surface elevation from BedMachine \cite{Morlighem2017BedMachineConservation} with a vertical exaggeration of 25. \textbf{B}, \textbf{C}, and \textbf{D} detail of plumes corresponding to colored arrows in \textbf{A} with data from \citeA{CReSIS2013RadiostratigraphyCenter}.}
\label{fig:plumes}
\end{figure}

At first glance, an entirely separate problem is the nature of the formation of large (\(>\)1/3 the local ice thickness) englacial plumes found by tracing reflections of equal age in radargrams (i.e., isochrones; Figs. \ref{fig:location}A, \ref{fig:plumes}, S1, \citeA{CReSIS2013RadiostratigraphyCenter}), which complicate reconstructions of palaeoclimate through layer-thickness inversions \cite{Theofilopoulos2023SensitivityDynamics, Rieckh2024DesignTracing}. Mapping \cite{LeysingerVieli2018BasalStratigraphy} and visual inspection of automatically-tracked disrupted radiostratigraphy \cite{Panton2015AutomatedSheet} shows that these large plume-like features (hereafter plumes) are mostly found in the northern part of the GrIS (Figs. \ref{fig:location}A, S1). Such plumes have previously been hypothesized to result from basal freeze on \cite{LeysingerVieli2018BasalStratigraphy}, or traveling basal slippery spots \cite{Wolovick2014TravelingSheets}, which both require an at least temporarily thawed bed. Separately, \citeA{Bons2016ConvergingSheet, Zhang2024FormationSheet} show that convergent flow, rheological anisotropy, a rough bed, and viscosity gradients are sufficient to form small-scale (\(<\)100 m) folds, but that large-scale folds require density gradients induced by thermal expansion and significantly lower viscosity. Another way to describe such temperature- and buoyancy-driven fold formation -- and which lies along the same process continuum -- is convection. In thermal convection, layers of ice heated geothermally from below (or cooled from above) thermally expand at the bottom (or thermally contract at the top), creating an unstable density gradient, forcing a vertical flow of material. Here we use ``local convection" to refer to a temperature- and density-controlled process generating self-sustaining upwards motion and disrupted plume-like structures emanating from the bed that are relatively isolated spatially. We explore whether ice convection -- a process with a contentious history in theoretical glaciology -- can explain observations of these large plumes, also known as disrupted basal units. We further consider the implications of convection potentially being the dominant process in their formation and whether the unique physical conditions that are needed for convection to operate can further constrain GrIS rheology.
%\citeA{Zhang2024FormationSheet} describe these as buoyancy enhanced folds, but the distinction between buoyancy enhanced folds and convection lies along a gradual natural continuum. We opt for the latter term as it is simpler and in keeping with its general usage to describe thermally-controlled changes in density that result in buoyant forces and subsequent motion. 

Convection has previously been proposed for ice sheets \cite{Hughes1976TheSheets, Hughes2012AreSheet}, but only for full-thickness convection (rather than stagnant-lid) and meeting firm objection \cite{Fowler2013ThermalSheets}. Both authors approach convection analytically only, by estimating a Rayleigh number \(Ra\), the dimensionless ratio of heat transfer via upwards mass transport (i.e. convection) vs. thermal conduction \cite{Rayleigh1916LIX.Side}. In these analytical models, convection initiates when a critical Rayleigh number is exceeded (\(\sim650-1700\) in \citeA{Knopoff1964TheHypothesis} and \citeA{Hughes1976TheSheets}), reached at \(E=5\) if the effective viscosity is assumed to be vertically uniform through the entire ice column; Supplementary Material). While these studies found \(Ra\) close enough to the critical value to warrant consideration of convection, applying a purely analytic approach to the GrIS is not ideal. The formulation of thermal diffusion in \(Ra\) does not capture dynamical effects important in ice sheet flow, such as horizontal shearing; and the critical \(Ra\) is itself tied to the particular boundary conditions and the initial perturbation geometry (e.g., \cite{Solomatov1995ScalingConvection}) making an analytical approach challenging for a system perched close to the onset of convection behavior. We therefore consider a direct numerical modeling approach as a more appropriate method of investigation for the question of convection in terrestrial ice sheets.

\section{Materials and Methods}
We use the geodynamics software package ASPECT 2.5.0 \cite{Kronbichler2012HighMethods, Heister2017HighProblems, Bangerth2023ASPECT2.5.0} with a setup adjusted to simulate a 25-km along-flow two-dimensional slice through an ice sheet (Fig. S3), or a 22 km along-flow by 18 km across-flow three-dimensional cuboid (Fig. S2). ASPECT is used in place of a conventional ice sheet model due to its extensive benchmarking in convection problems and in-built functionality to handle buoyancy forces, which are lacking in modern ice-sheet models. Similar geodynamics models have been previously used to study convection in the shells of icy moons \cite{Lebec2023ScalingHabitability}. To facilitate a broad parameter sweep at low computational expense, and to isolate the influence of the parameters in question, we simplify the domain to have a uniform ice thickness (2.5 km as a default, though we explore other choices). Surface mass balance is set to zero, i.e., no snowfall or surface melting (\(v_{z,s}=0\) at the surface boundary condition where \(\boldsymbol{v}= (v_x, v_y, v_z)\) is the velocity field, the subscript \(x\) represents the along-flow distance, the subscript \(z\) is depth, and the subscript \(y\) is across-flow distance in 3-D simulations) for all simulations except those that explicitly consider snowfall. When applied, surface shearing velocity \(v_{x,s}\) is uniform across the domain's top surface with the rigid \(v_{x,b}=0\) condition maintained at the base. Keeping \(v_{x,b}=0\) is likely a firmer control on basal velocity than the possible décollement observed in radargrams (Fig. \ref{fig:plumes}C), i.e., while increasing \(v_{x,s}\) in the model can simulate plume behavior as \textit{actual} surface velocity increases (Fig. \ref{fig:location}C), the comparison is not one-to-one. Similarly, the placement of the initial perturbation in snowfall runs will influence the balance between horizontal and vertical velocity components (Fig. S5). Snowfall, surface velocity, and ice thickness all furthermore exhibit moderate variation over the millennial timescales important for convection \cite{MacGregor2016HoloceneSheet}. Resolution is determined by the ASPECT requirement to set grid spacings as a given number of even divisions. In the case of a 2,500 m thickness and 6 divisions this gives horizontal and vertical resolutions of 390 and 39 m, respectively.

In most ice-sheet models, stress is related to strain with the Nye-Glen isotropic flow law \cite{Nye1952TheFlow, Glen1955TheIce}
\begin{equation}
    \dot{\boldsymbol{\epsilon}}=A\tau_{e}^{n-1}\boldsymbol{\tau}
    \label{eq:glen}
\end{equation}
where \(\dot{\boldsymbol{\epsilon}}_{ij} = \frac{1}{2}\left(\frac{\delta v_i}{\delta x_j} + \frac{\delta v_j}{\delta x_i}\right)\) is the strain rate tensor, \(\tau_{e}^{2}=\frac{1}{2}\mathrm{tr}(\boldsymbol{\tau}^{2})\) (Pa) is the effective stress and \(\boldsymbol{\tau}\) is the stress tensor, \(n\) is the flow exponent generally assumed as 3 or 4 \cite{Cuffey2010TheGlaciers, Bons2018GreenlandMotion}, and \(A = E A_0\mathrm{exp}\left( -\frac{Q}{R}\left(\frac{1}{T}-\frac{1}{T_0}\right)\right)\) (Pa\textsuperscript{-\(n\)} a\textsuperscript{-1}) is the creep parameter, where \(E\) is the enhancement factor, \(A_0\) is the creep prefactor, \(Q\)  (J mol\textsuperscript{-1}) is the activation energy, \(R\) is the ideal gas constant (J mol\textsuperscript{-1} K\textsuperscript{-1}), \(T_0\)= 263.2 K, and \(T\) (K) is ice temperature. The influence of \(E\) on rheology will vary depending on the choice of \(A_0\) and \(n\); we use the default values in \citeA{Cuffey2010TheGlaciers} as a widely used reference (Table S1) that makes comparison with existing ice-sheet models more straightforward. In those models, separate \(Q\) values are used for high (\(T>T_0 \approx\) 263.1 K) and low temperatures, but for simplicity in ASPECT here we simplify this to one mid-range value, which has limited effect (Fig. S6). Moreover, the pressure dependence of $A$ is neglected. We also simplify to a Newtonian rheology for ASPECT by setting \(\tau_{e}^{n-1}\) as constant with \(\tau_e = 50\,\)kPa and \(n=3\), meaning that the temperature-dependent viscosity is then controlled by varying \(E\) (Fig. S3). A Newtonian rheology is appropriate here as strain rates due to convection are small compared to those from background ice flow (Fig. S5), and convection can be considered a secondary phenomenon in this sense. Prescribing \(\tau_e\) is also necessary as a full ice-sheet stress state can not be accurately replicated in a simplified along-flow slice. The \(\tau_{e}\) value (Fig. \ref{fig:location}D) is taken from an Ice-sheet and Sea-level System Model simulation \cite{Larour2012ContinentalISSM} (Supplementary Material). Effective viscosity \(\eta\) can then be calculated as
\begin{equation}\label{eq:etadef}
\eta = \frac{1}{2}\left[A\tau_{e}^{n-1}\right]^{-1} .
\end{equation}
Rearranging Eq. \ref{eq:etadef} yields the functional dependence of $E$:
\begin{equation}\label{eq:Efother}
E = c\eta^{-1}\tau_e^{1-n},
\end{equation}
where $c = (2A_0)^{-1}\exp\left(\frac{Q}{R}\left[T^{-1}-T_0^{-1}\right]\right)$. 

ASPECT solves the governing equations of convection,
\begin{alignat}{2}\label{eq:convection}
    % \begin{split} 
    \nabla \cdot \boldsymbol{v} &= 0 && \qquad \text{(conservation of mass)} \\
    - \nabla \cdot \left[ 2 \eta \dot{\boldsymbol{\epsilon}} \right] + \nabla p^\prime &= -\beta \bar{\rho} T^\prime g \hphantom{_{\rm int}} && \qquad \text{(conservation of momentum)} \\
    \bar{\rho} C_p \left( \frac{\partial T}{\partial t} + \boldsymbol{v} \cdot \nabla T \right) - \nabla \cdot \kappa \nabla T &= F_{\rm int} && \qquad \text{(conservation of energy)}
    % \end{split}
    %\mathrm{tr}(\dot{\boldsymbol{\epsilon}}\boldsymbol{\tau})
    %\bar{\rho} C_p \left( \frac{\partial T}{\partial t} + \boldsymbol{v} \cdot \nabla T \right) - \nabla \cdot \kappa \nabla T &= F_{\rm int} 
\end{alignat}
where $p$ (Pa) is pressure, $C_p$ (J\,kg$^{-1}$\,K$^{-1}$) is heat capacity, $\kappa$ (W\,m$^{-1}$\,K$^{-1}$) is thermal conductivity, \(\beta\) (K$^{-1}$) is thermal expansion coefficient, \(\bar{\rho}\) (kg\,m$^{-3}$) is the reference density, \(g\) (m\,s\textsuperscript{-2}) is acceleration due to gravity, and \(F_{\rm int}\) (W m\textsuperscript{-2}) is the sum of all other heating terms. We set \(F_{\rm int}=0\), thereby ignoring adiabatic heating and neglecting strain heating, to prevent simulations with greater \(v_{x,s}\) and hence greater strain heating from evolving a different rheology along flow. These equations follow the Boussinesq approximation -- that density variations are small enough to be neglected everywhere except for in the buoyancy term $\beta \bar{\rho} T^\prime g$ -- which is valid for very slow-flowing materials without abrupt density changes. This solution method simplifies the temperature field to $T = \bar{T} + T^\prime$, where $\bar{T}$ is a constant reference temperature and $T^\prime$ is the temperature perturbation; analogous perturbations are formed for the pressure and density fields.

Two baseline temperature profiles are used, representing the colder NEEM ice-core site in northern Greenland \cite{Dahl-Jensen2013EemianCore, Rasmussen2013ACore} and the warmer DYE-3 in southern Greenland \cite{Gundestrup1984Bore-HoleGreenland} respectively (Figs. \ref{fig:location}A, S3). We apply a transformation, \(T_2 = \frac{T_1 + T_a}{T_b}(T_b + T_a)\) where \(T_1\) is the original temperature profile, \(T_b\) is the basal temperature and \(T_a\) is an adjustment term used to raise the basal temperature to $-2^{o}$C. The temperature profiles are stretched and compressed when adapted to the range of ice thicknesses. As we keep the basal temperature uniform across the domain and also want to consider ice some distance from the ice core site, a slightly higher fixed basal value is appropriate as a midpoint between the ice-sheet interior and margins. An initial temperature perturbation replicating a fold is created 5 km in from the inflow side (3.5 km in the case of snowfall simulations) using two Gaussian functions of opposing signs. We refer to two temperature perturbation sizes for 2D runs: large, used for most simulations (Fig. S3), and medium (Fig. S4). In the 3D runs a simpler approach is taken, with a cube of uniform 273 K ice measuring 3,000 m \(\times\) 5,000 m \(\times\) 750 m as the initial perturbation. As the initial temperature gradient is not linear (Fig. S3), using a larger initial perturbation allows convection to occur in a more realistic temperature field without a delay for initial plume development, though results are relatively insensitive to the initial perturbation (Figs \ref{fig:results}B, \ref{fig:results}C, S7). ASPECT input files and scripts to recreate the temperature perturbations and perform other post-processing operations are provided in the Open Research Section. Values for set parameters are given in Table S1. 

For each temperature profile we focus on the influence of four variables on the maximum upwards-directed vertical velocity, \(\text{max}(v_z)\): the enhancement factor (\(E\)), shear velocity (\(v_{x,s}\)), ice thickness (\(H\)), and snow accumulation rate (\(v_{z,s}\)) (labeled in Table S2, Fig. \ref{fig:results}). Additional 3D simulations are included for runs B and F which consider the parameter space covering observed plumes (NEEM temperature profile). Defining a threshold for convection under a given parameter space is not straightforward, but we focus on \(\text{max}(v_z)\) over time as a reasonable indicator. Nonetheless, even if \(\text{max}(v_z)\) trends towards zero over time, the englacial stratigraphy will still be slightly disrupted during this transition period. The total buoyancy forces in the 3D simulation will also be greater than in 2D as we are able to model an isolated plume rather than a laterally extensive fold. We divide behavior into three zones, focusing on the 2D simulations that cover a broader parameter space. \textbf{Suppressed} convection is defined where \(\text{max}(v_z)\) at 20 kyr is below 0.01 m yr\textsuperscript{-1} or where \(\text{max}(v_z)\) at 20 kyr has dropped by 0.03 m yr\textsuperscript{-1} or more relative to its value at 4 kyr. \textbf{Amplifying} convection is defined where \(\text{max}(v_z)\) at 20 ka exceeds 0.4 m yr\textsuperscript{-1} or where \(\text{max}(v_z)\) has increased by 0.1 m yr\textsuperscript{-1} or more between 4--20 kyr ka. \textbf{Sustained} convection then occupies the space between these two zones. This approach allows us to isolate which parameter combinations may produce sufficient upwards flow to account for the distribution of large englacial plumes (Fig. \ref{fig:location}A).

\section{Results}
\begin{figure}
\noindent\includegraphics[width=1\textwidth]{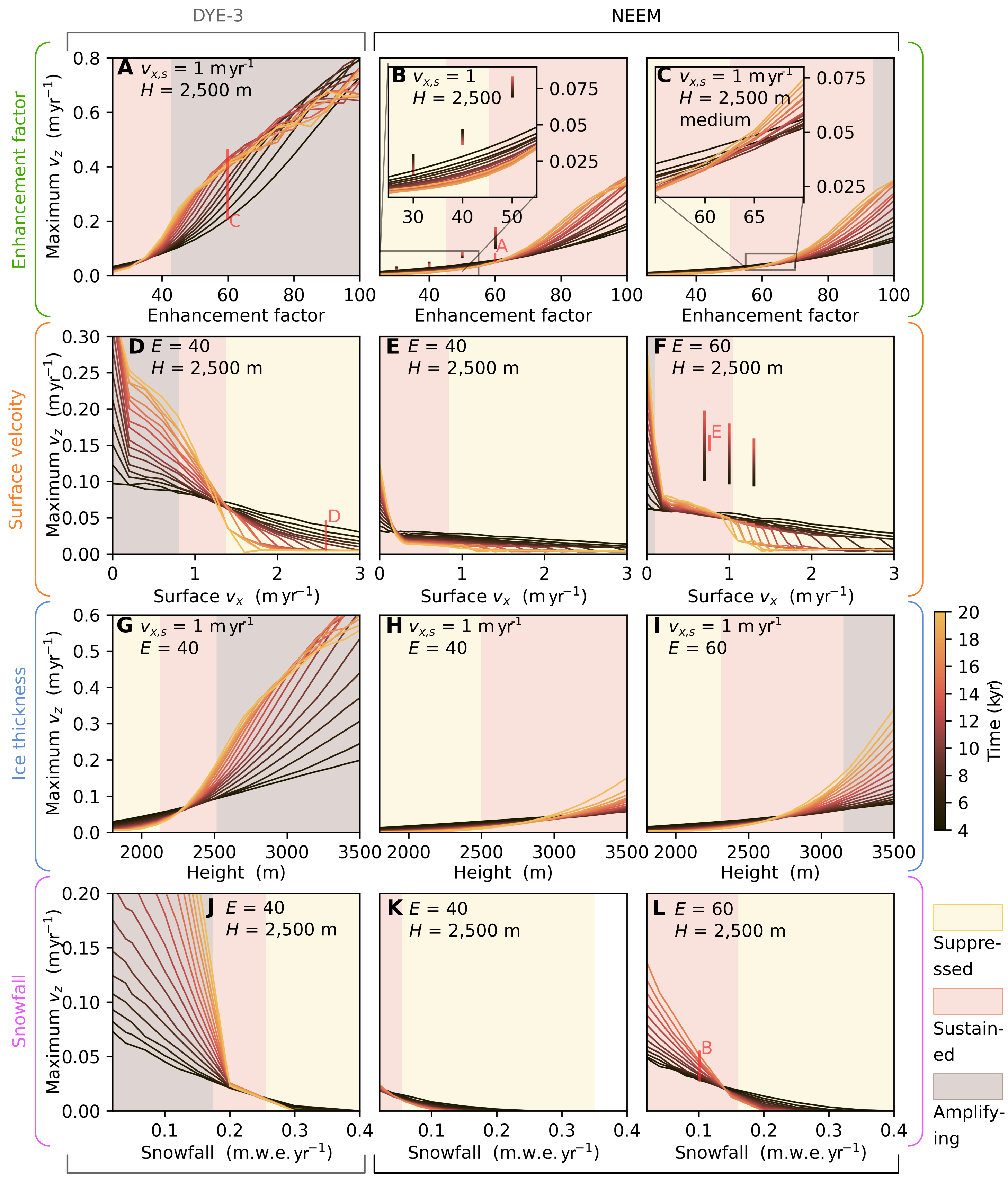}
\caption{Convection model ensemble results. Relevant model parameters are given in the top left of each panel and along the left and top figure border. All runs use the large initial perturbation except for \textbf{C}. Red lines and letters in \textbf{A}, \textbf{B}, \textbf{D}, and \textbf{F} refer to the runs shown in the correspondingly labeled panels of Fig \ref{fig:plumes_out} and vertical lines in \textbf{B} and \textbf{F} correspond to 3D simulations. All simulations except for those with non-zero snowfall (\textbf{J}, \textbf{K}, \textbf{L}) have \(v_{z,s}=0\).}
\label{fig:results}
\end{figure}

\begin{figure}
\noindent\includegraphics[width=0.8\textwidth]{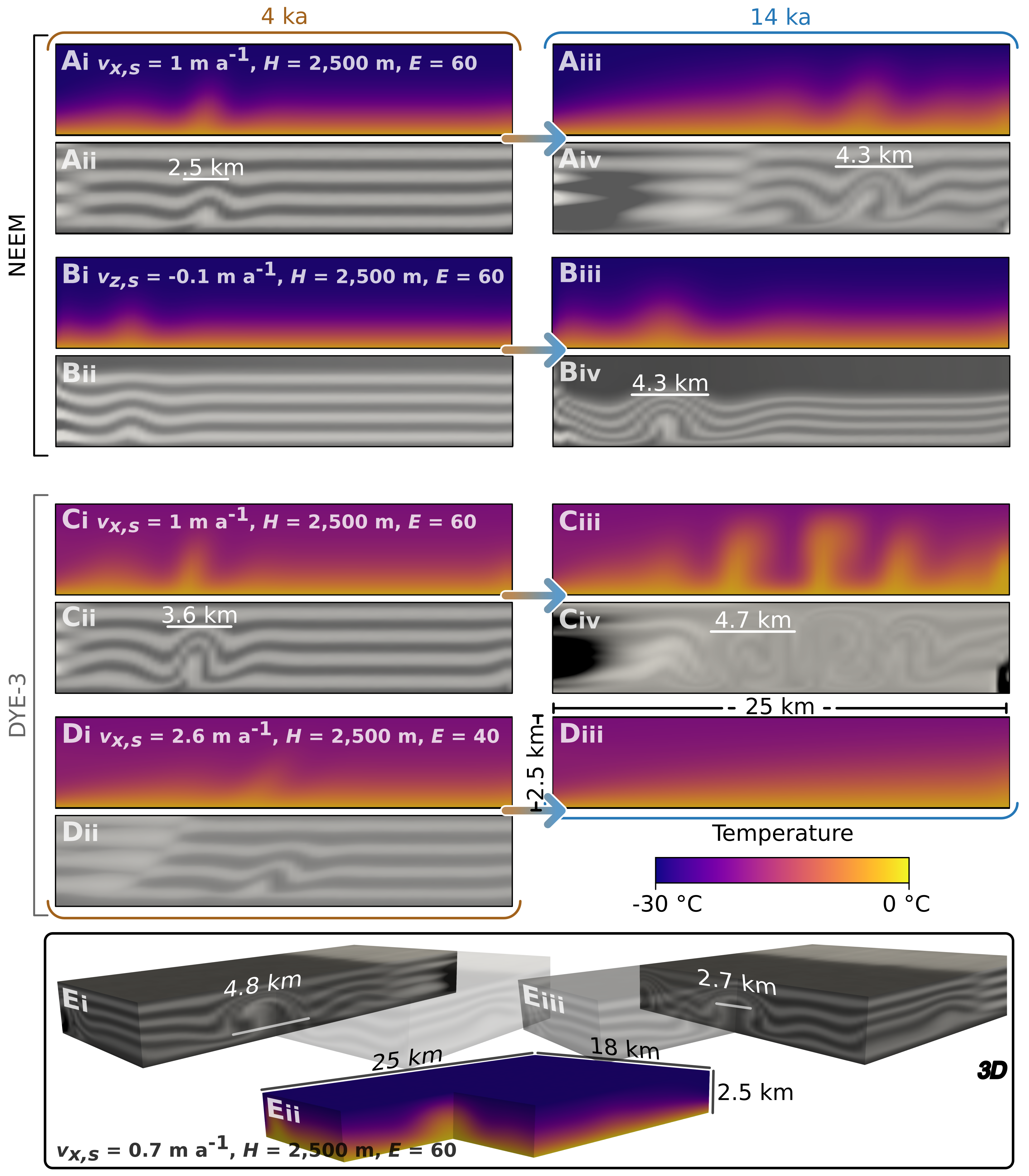}
\caption{Snapshots of model behavior at 4 and 14 kyr as marked in Fig. \ref{fig:results}. Panels with plasma colormap display temperature, while the accompanying grayscale colormap in each instance shows disruption of initially flat horizontal layering used to simulate isochrones. Numerical diffusion leads to the pattern in \textbf{Civ} being slightly difficult to distinguish. The pattern corresponding to \textbf{Div} is unintelligible and therefore omitted. Each domain has a length of 25 km, a height of 2,500 m and a vertical enhancement factor of 2. Panel \textbf{E} shows the 3D simulation run corresponding to the marking in \ref{fig:results}F. Fig. S5 mirrors Fig. \ref{fig:plumes} but displays \(v_x\) and \(v_z\).}
\label{fig:plumes_out}
\end{figure}

While our modeling is substantially more sophisticated than calculation of a single \(Ra\) value, the general behavior in our simulations can still be understood in terms of the \(Ra\) number (Supplementary Material), with greater values of \(E\) and \(H\) prompting convection. However, given the additional complications of varying surface velocity, snowfall, initial perturbation, and viscosity profile, we find that there is no single critical value of \(E\) that describes this transition, but Fig. \ref{fig:results} suggests that \(45\leq E\leq75\) encapsulates a range of behavior for the NEEM temperature profile sufficient to form features similar to those observed in radiostratigraphy (Fig. \ref{fig:plumes_out}). Considering shear over the domain of 1 m yr\textsuperscript{-1} with no snowfall (Fig. \ref{fig:results}B) in 3D, \(\text{max}(v_z)\) begins to increase from 4--14 kyr between \(40<E<50\), around the same point at which 2D convection is considered to be sustained under our definition. In Fig. \ref{fig:results}B at \(E=75\), \(\text{max}(v_z)\) is consistently increasing over time and when \(E=60\) convection is still classified as sustained for snowfall rates exceeding 0.15 m yr\textsuperscript{-1}.

DYE-3 requires much lower \(E\) values to transition between suppressed, sustained, and amplifying zones compared to NEEM when other parameters are equivalent (cf. Figs. \ref{fig:results}A, B and Figs. \ref{fig:plumes_out}A, C). The steep temperature gradient at the base of the DYE-3 profile is sustained over a shorter height, resulting in lower buoyancy forces overall, but this is compensated by lower viscosity in the upper portion of the domain (Fig. S3). Ice thickness is an important factor for both profiles, with sustained convection becoming infeasible below thicknesses of around 2,000 m (hence its use as a boundary in Fig. \ref{fig:location}B), though note that ice thickness is also intertwined to an extent with its influence on the basal temperature gradient. Vertical transport rates begin to level off with increasing ice thickness for the DYE-3 profile, pointing towards a maximum rate of upwards motion. However, surface velocity and snowfall exert perhaps the most important overall constraints on \(\text{max}(v_z)\), with convection becoming infeasible as surface velocity in our simulations increases from 1 to 3 m yr\textsuperscript{-1} or as snowfall increases beyond 0.1 to 0.3 m yr\textsuperscript{-1}, with the precise cut-off value depending on \(E\) and the temperature profile (Figs. \ref{fig:results}D-F, J-L). %Note again however that while these results provide a good indication of convection behavior with increasing surface velocity and snowfall, they do not definitively indicate an exact relationship for ice-sheets that may have more complicated velocity fields and boundary conditions. %where the lowermost part of the ice column where plumes are not observed may accommodate significant shear (Fig. \ref{fig:plumes}. %You: I think this whole section can be cut back/rephrased more simply
%Dec 31, 2024 3:30 PM • 
% • %Claire Marie Guimond: I agree this paragraph is sort of confusing. I think Fig 3 could be pointed to upfront as showing how both (1) max(v_z) and (2) rate of change of max(v_z) with time depend on the various parameters
%Jan 12, 2025 4:35 PM %Claire Marie Guimond: as it currently stands, this is focusing on constraining E first, before just describing the overall behaviour, which maybe seems like jumping ahead

The modeled plume geometry varies substantially across the parameter space. In the suppressed convection zone, a perturbation is still produced but is not sustained (Figs. \ref{fig:results}D, \ref{fig:plumes_out}D). Amplifying convection can prompt a self-sustaining plume chain with significant temperature variation (Figs. \ref{fig:results}A, \ref{fig:plumes_out}C). Sustained convection (Figs. \ref{fig:results}B, F, L, \ref{fig:plumes_out}A, B, E) produces plumes more similar to the folds observed in radargrams, though plume length is slightly shorter. 3D simulations produce roll-over in the down-flow direction, but characteristic symmetric overturning typical of convection in an otherwise static medium in the across-flow direction (Figs. \ref{fig:plumes}D, \ref{fig:plumes_out}E). Using a smaller perturbation in 2D simulations does not appreciably alter this pattern; instead it merely shifts the time at which each \(\text{max}(v_z)\) is reached (Fig. \ref{fig:results}C). The heat flux required to sustain the fixed basal temperature (\(\sim\)40-70 mW m\textsuperscript{-1}, Fig. S5) is compatible with proposed rates of geothermal heat flux beneath the GrIS \cite{Zhang2024EvaluatingSheet}.

\section{Discussion}

Our results indicate that four main thresholds must be reached for convection to occur within the Greenland Ice Sheet: (1) Ice thickness must be greater than around 2,200 m; (2) Total horizontal shear through the column must be less than around 1 m yr\textsuperscript{-1}; (3) Snowfall must be less than around 0.15 m yr\textsuperscript{-1}; and (4) the enhancement factor must exceed around 45-75. Condition (1) is satisfied for a large region of the interior of the central ice sheet  (Fig. \ref{fig:location}B), condition (2) is likely satisfied by low surface velocities throughout northern central Greenland (Fig. \ref{fig:location}C), with longer residence time being unique to the northern central ice divide (Fig. \ref{fig:location}B), and condition (3) is also primarily satisfied in north Greenland \cite{MacGregor2016HoloceneSheet}. %main factors are conducive to convection The parameter space in which convection can occur within the GrIS (Results) suggests that the mechanism falls within the realm of possibility. Plumes in north Greenland fall within a ``local convection" morphology; i.e., observed plumes are relatively isolated and closer in style to Figs. \ref{fig:plumes_out}B, D than Fig. \ref{fig:plumes_out}A, and reproduced with relatively good skill when \(40 \leq E\ \leq 60\). 

We next consider the possibility that \(E\) is 9--15 times larger than commonly assumed for north Greenland \cite{Cuffey2010TheGlaciers}, satisfying condition (4), after first considering other aspects of plume morphology and distribution. First, modeled plume widths (Fig. \ref{fig:plumes_out}) are comparable but slightly narrower than observations. We suggest this may occur due to continued disruption of plumes as the velocity field evolves after they have attained their maximum amplitude. Notably, the convection plumes generated in our 3D simulations (Fig. \ref{fig:plumes_out}E) more closely resemble the geometry of observed folds (Fig. \ref{fig:plumes}C, D), and we are not aware of another mechanism that is hypothesized to produce this unique type of geometry. Second, large plumes are found outside of the region bounded by the principal 2,000 m contour (Fig. \ref{fig:location}B) and in regions of fast-flow (Fig. \ref{fig:location}). However, as we assume steady-state formation conditions and our plumes have a formation time of $\sim$14 kyr, if our convection hypothesis is correct then these plumes would have initiated when the GrIS was thicker, colder and larger \cite{Lecavalier2014AExtent}. Large plumes found closer towards the ice sheet margins (Fig. \ref{fig:location}A) may therefore have formed initially farther inland where surface velocity is lower (Fig. \ref{fig:location}C), before being advected while retaining their general morphology. Indeed, all plume clusters begin well into the ice-sheet interior though we note that as effective stress increases towards the onset of fast-flowing regions, effective viscosity will also decrease (Fig. \ref{fig:location}D). Such settings are rarer in southern Greenland, where escape times from the central ice divide to the 2,000 m principal contour line are only a little over 10 kyr. Third, the significant snowfall (\(\geq\) 0.35 m yr\textsuperscript{-1}) and hence downwards motion in south Greenland (Fig. \ref{fig:location}E) likely limits the possibility of convection in this region (Fig. \ref{fig:results}J) and will have done for at least the past 9 kyr \cite{MacGregor2016HoloceneSheet}.

We next consider existing hypotheses for formation of the observed folds. Basal freeze-on may be limited as a general explanation capable of explaining plume ubiquity in north Greenland \cite{Bell2014DeformationMeltwater, LeysingerVieli2018BasalStratigraphy} given the requirements for large volumes of basal water \cite{Dow2018LimitedStratigraphy} in a region that is not likely to be pervasively thawed \cite{Bons2018GreenlandMotion, Macgregor2022GBaTSv2:Sheet}. Freeze-on may furthermore not explain the fairly consistent sizing and spacing (at \(\sim\)10 km, Fig. \ref{fig:plumes}) of north Greenland plumes, which contrast the much more spatially extensive freeze-on layers in East Antarctica \cite{Bell2011WidespreadBase}. Traveling slippery spots \cite{Wolovick2014TravelingSheets} develop clearly in an controlled setup, but also require thawed bed areas in the same region and further do not appear to align with the observation of a highly deformed basal layer beneath the plumes (Fig. \ref{fig:plumes}C), which may be more consistent with high rates of basal ice deformation than basal sliding \cite{Zhang2024FormationSheet}. Traveling slippery spots may also not be compatible with the 3D geometry of observed plumes (Figs. \ref{fig:plumes}B, D) or with ice motion over a rough bed. Additionally, neither mechanism accounts for an apparent absence of \(H>1/3\) plumes in south Greenland. However, the basal thermal state may have been different \(\sim\)10 kyr ago and we cannot rule out these two processes contributing to the onset of an initial perturbation. We highlight these possibilities to motivate further work on englacial plumes to more clearly determine if convection is indeed a primary mechanism for their formation.
%Traveling slippy spots also produce plumes that can extend 10s of km in the along flow direction, out of keeping with their observed morphology

Are \(E\) values 9-15 times larger than commonly assumed feasible for the northern GrIS? Independent of convection being possibly the only feasible mechanism for large plume formation, we suggest that an affirmative answer is appropriate. Large plumes are mostly found in areas with a relatively larger proportion of pre-Holocene ice (Fig. \ref{fig:location}A). Beyond this observation fulfilling the requirement for relatively stable ice (condition (2)), older ice from the Last Glacial Period is consistently measured or inferred to be significantly less viscous than Holocene ice \cite{Paterson1991Whysoft, MacGregor2016HoloceneSheet, Bons2018GreenlandMotion, Law2021ThermodynamicsSensing}, as a result of fabric development and a higher impurity content. Despite the importance of this softer ice for interpretation of the overall motion of the GrIS, very few direct measurements exist. Borehole closure rates from ice divides likely reflect stresses inconsistent with basal shearing, and such locations are often explicitly selected for their lack of a history of extensive horizontal shear \cite{Talalay2007ClosureDiscussion}. To our knowledge, no laboratory measurements have been conducted on ice resembling that found within what we hypothesize to be englacial convection plumes. It may therefore be possible that basal ice in north Greenland is sufficiently soft as to permit convective plume formation. Tests on field specimens present the clearest opportunity to directly asses our hypothesis.

In situ rheology is also modulated by anisotropy, which is not included in our simulations, although the enhancement factor is sometimes used in ice sheet models to account for anisotropy in a simple manner and functions similarly in our model \cite{Cuffey2010TheGlaciers}. \citeA{Zhang2024FormationSheet} suggest an important role for anisotropy in the formation of large plumes (their Fig. 4). However, it is possible that this is a result of their implementation of anisotropy, which enables the viscosity acting parallel to the plane perpendicular to the $c$-axis direction to decrease by a factor of three and fall below the \(1\times10^{13}\) Pa limit set for the isotropic run (their Table S2). Notably, this decrease is sufficient to reach the effective basal viscosity values (\(\sim 3\times10^{12}\) Pa s) in our \(E\)=40 and \(E\)=60 simulations, where local convection becomes increasingly viable. Ice may also be more non-linear than its implementation in this study, with growing evidence for \(n=4\) in some regions \cite{Bons2018GreenlandMotion, Ranganathan2024ASheets}. Our use of Newtonian rheology allows us to relate \(E\) to \(n\) (Eq. \ref{eq:Efother}), and to use a commonly implemented consensus value for \(A_0\) \cite{Cuffey2010TheGlaciers}, but we anticipate that \(n=4\) will slightly alter \(E\) values, and may increase the importance of the non-linear stress response within the plumes. We emphasize, however, that rate-weakening in plumes is still anticipated to be small compared to the main coastward movement of the ice sheet which exerts a first order control over effective stress (Figs. \ref{fig:location}D, S1, S3).

A lower effective viscosity of basal ice will significantly influence ice dynamics, similar to the influence of an increased flow exponent, \(n\) \cite{Bons2018GreenlandMotion, Zeitz2020SensitivityGeometry, Ranganathan2024ASheets}. If ice-sheet models are initiated under fixed assumptions of higher ice viscosity, then inversions for basal traction will overcompensate by producing unrealistically low basal traction values and bias the resulting projections \cite{Berends2023CompensatingResponse}. Convection-driven plumes also present a mechanism that draws warmer and lower-viscosity basal ice upwards -- counteracted by the transport of higher-viscosity (colder) interior ice downwards. Exploring the possible implications of lower viscosity ice and convection-driven mixing -- and their influence upon inferred basal traction -- is therefore warranted to better quantify the errors that may be introduced into predictive ice-sheet models. Finally, the relative lack of large plume observations in the Antarctic Ice Sheet, outside of the Gamburtsev Mountains, may simply result from colder temperatures there and hence higher viscosities in the upper ice column, which limit convection \cite{Fortuin1990ParameterizationAntarctica, Bell2011WidespreadBase, Cavitte2021AMid-Pleistocene, Sanderson2023EnglacialAntarctica}, or from a sampling bias given the comparative paucity of radar-sounding observations in Antarctica \cite{Bingham2024ReviewSheets}. 

\section{Conclusions}

Our modeling indicates that local convection is possible within the Greenland Ice Sheet under conditions that are not unrealistically far from the existing consensus on ice rheology. This hypothesis could explain the observed spatial distribution of large plumes in Greenland, with surface velocity, accumulation rates and ice rheology exerting the strongest controls on convection viability and hence plume formation. A direct corollary of this result is that ice in northern Greenland may be 9-15 times softer than commonly assumed. Appropriately implementing these constraints into ice-sheet models may help reduce compensatory errors and improve the accuracy of their future projections. 

\section*{Open Research Section}
Radar flight line data is available from \citeA{Macgregor2015RadarSheet}. The ASPECT and ISSM input files and scripts used to obtain results are available from (Harvard Dataverse repository to be created). Plume location data comes from \citeA{LeysingerVieli2018BasalStratigraphy}.

\acknowledgments
RL, AB, and PV acknowledge funding from Norges Forskingsråd (SINERGIS project, Norwegian Research Council Grant 314614). CMG is supported by the UK Science and Technology Facilities Council [grant number ST/W000903/1]. JM acknowledges support from the NASA Cryospheric Sciences Program. Thanks to Gwendolyn Leysinger Vieli for the background information on the plume locations. 

\newpage

\appendix

\renewcommand{\thefigure}{S\arabic{figure}}
\renewcommand{\theequation}{S\arabic{equation}}
\renewcommand{\thetable}{S\arabic{table}}

\section{Supplementary Material}

\subsection{Rayleigh number}
The Rayleigh number following \citeA{Rayleigh1916LIX.Side} is calculated as

\begin{equation}
    Ra = \frac{H^3 \Delta T \beta g\rho}{ \alpha \eta},
    \label{eq:Rayleigh}
\end{equation}

where \(H\) (m) is the thickness of the fluid layer, \(\Delta T\) (K) is the temperature difference between the surface and base, \(\beta\) (K\textsuperscript{-1}) is the thermal expansion coefficient, \(g\) (m s\textsuperscript{-2}) is the acceleration due to gravity, \(\bar{\rho}\) is the base material density, and \(\alpha\) (m\textsuperscript{2} s\textsuperscript{-1}) is the thermal diffusivity. Previous attempts have been made to determine a critical Rayleigh number for non-Newtonian fluid layers (e.g. \citeA{Ozoe1972HydrodynamicSolution}) but this becomes complicated by their dependence on the amplitude and shape of the disturbance initiating motion \cite{Parmentier1978AFluids}.

If we extend the basal viscosity (calculated at 5\(\times\)10\textsuperscript{4} Pa effective stress and $-$2\degree C) uniformly through a 2,500 m ice column with \(\Delta T = 30\)K, we obtain a Rayleigh number of 2805 for \(E = 5\) (Fig. S3). \citeA{Hughes2012AreSheet} and \citeA{Fowler2013ThermalSheets} extend this approach in their arguments, with \cite{Fowler2013ThermalSheets} emphasizing that the lack of an initial thermal perturbation will prevent convection onset. However, bedrock perturbations (e.g. Figs. \ref{fig:plumes}B, C) or basal folding induced by other processes \cite{Zhang2024FormationSheet} can easily satisfy this challenge.

\subsection{ISSM run}
The ISSM (Ice-sheet and Sea-level System Model) run was completed following the setup of the UCI\_JPL group featured in \citeA{Goelzer2020TheISMIP6} with a higher-order Stokes approximation and an ISSM Budd sliding relationship relating basal traction, \(\tau_b\), to basal velocity, \(v_b\), as

\begin{equation}
    \tau_b = C^2 N^r v_{b}^{s}
\end{equation}

where \(C\) is the traction coefficient, \(r=\frac{q}{p}\), and \(s=\frac{1}{p}\) where \(q\) and \(p\) are parameters both set to 1. Following an inversion procedure to calculate basal traction the effective pressure, \(N\), is calculated as 

\begin{equation}
    N = \rho_i g H + \rho_w g b_z 
\end{equation}

where \(\rho_i\) and \(\rho_w\) are the densities of ice and water respectively and \(b_z\) is the position of the bed. \(n\) in Eq. 1 was set to 3, and an initial approximation of ice rigidity is made based on ice temperature. The model was run transiently for 0.25 years with a timestep of 0.01 a. The Matlab runscript can be found in the Open Research Section. This model was used only to give an indication of expected effective stresses within the GrIS and will reflect effective stress in most standard ice-sheet modeling applications. We did not re-run the model with updated enhancement factors suggested in this paper, or a greater value of \(n\), both of which may influence the effective stress and therefore effective viscosity. The reader is referred to \citeA{Larour2012ContinentalISSM} for a more detailed model and ice physics description.

\subsection{Model run times}

Each set of roughly 15 values each over a 20 ka period in 2D takes around 36 hr on 8 2-GHz CPUs. Running one 3D simulation for 14 ka takes around 84 hr on 48 2-GHz CPUs. 

\begin{figure}
\noindent\includegraphics[width=0.7\textwidth]{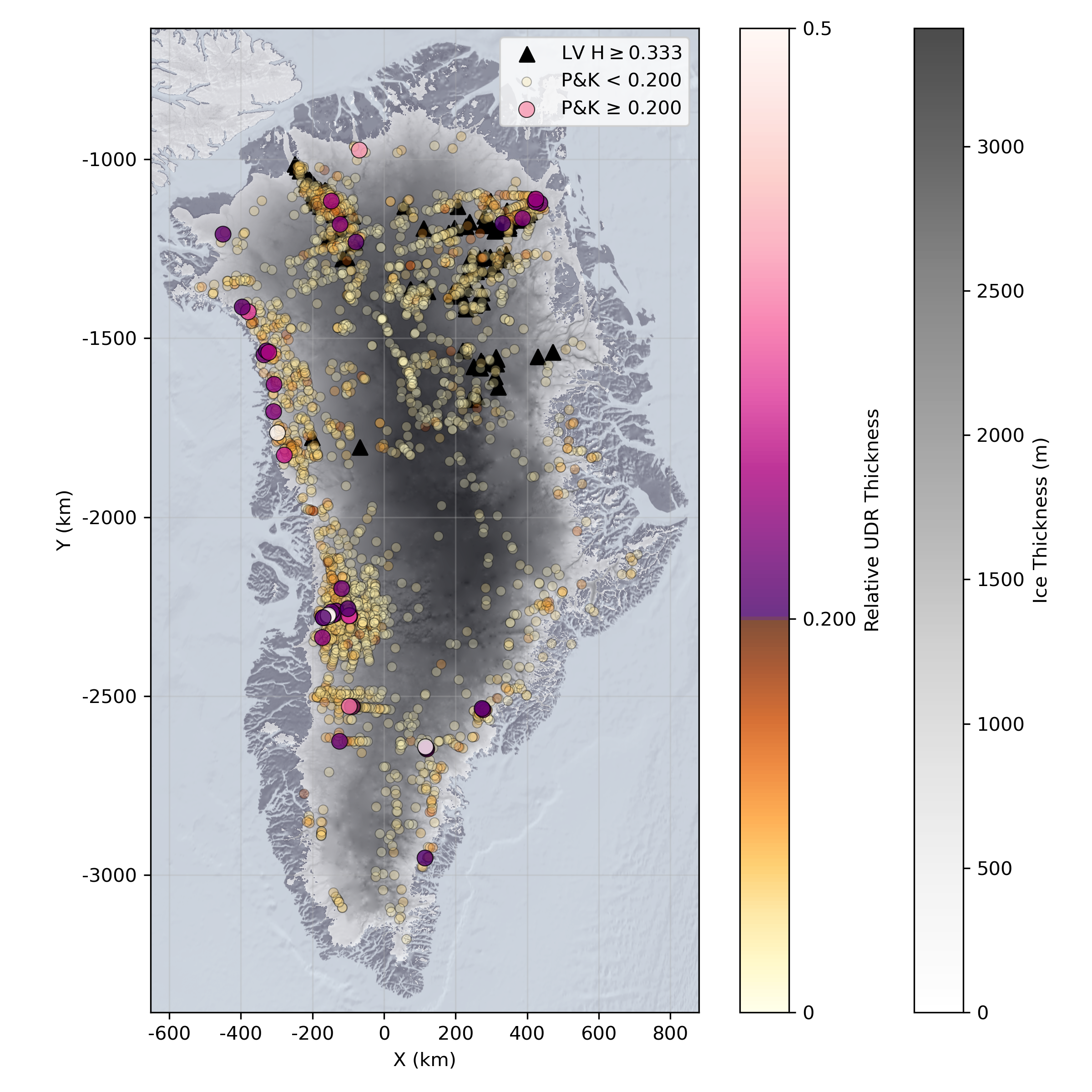}
\caption{Map showing automated mapping of units of disrupted radiostratigraphy (UDRs) from \citeA{Panton2015AutomatedSheet} (circles) and traced plumes from \citeA{LeysingerVieli2018BasalStratigraphy} (triangles). \citeA{Panton2015AutomatedSheet} UDRs are filtered to only include those detected where ice thickness exceeds 1 km. Relative UDR thickness is calculated using BedMachine v5 \cite{Morlighem2022IceBridge5}.}
\label{fig:}
\end{figure}

\begin{figure*}
\noindent\includegraphics[width=0.6\textwidth]{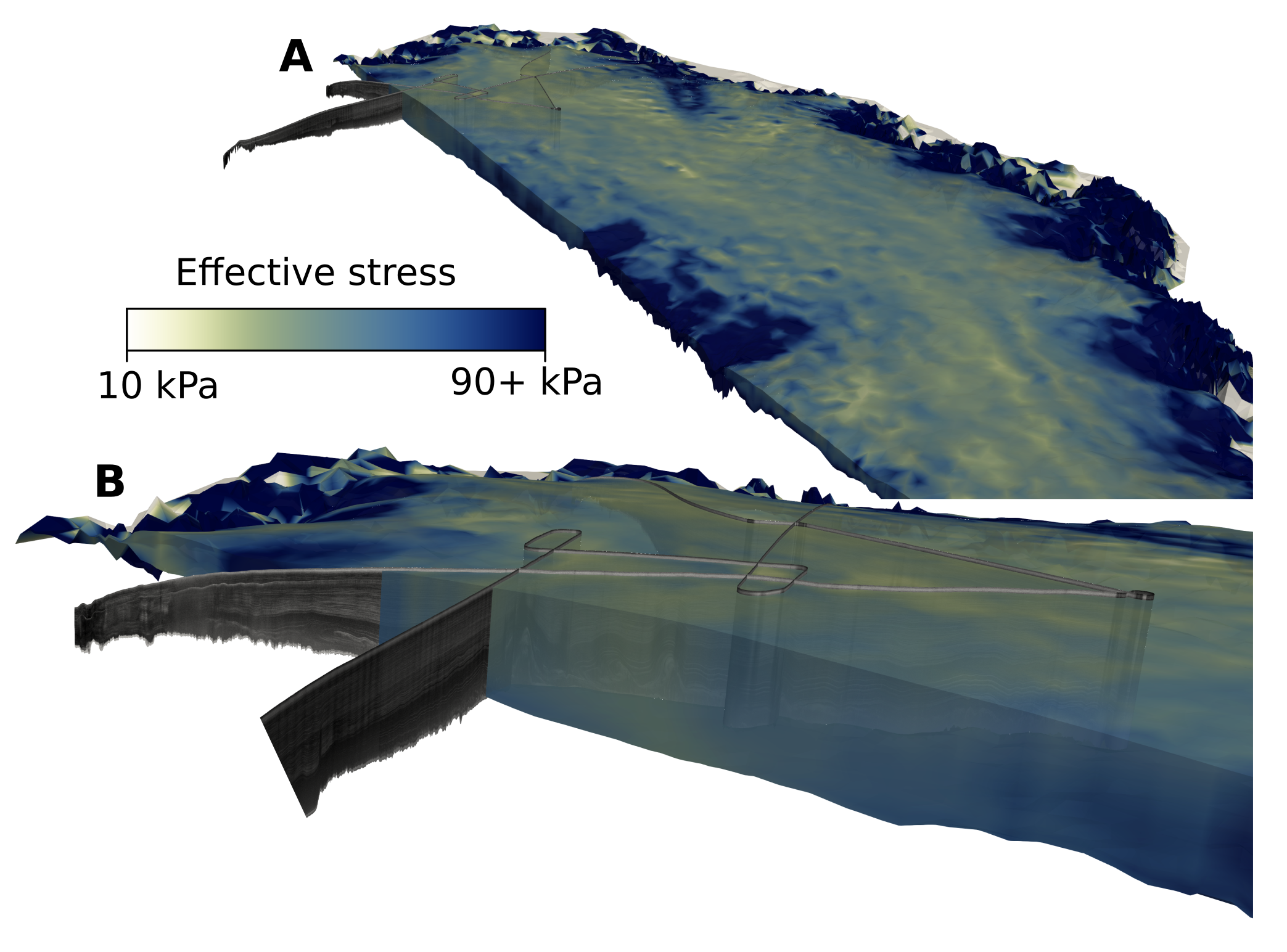}
\caption{Effective stress obtained from ISSM run with the same radar transects shown in Fig. 2.}
\label{fig:tau_e}
\end{figure*}

\begin{figure*}
\noindent\includegraphics[width=0.6\textwidth]{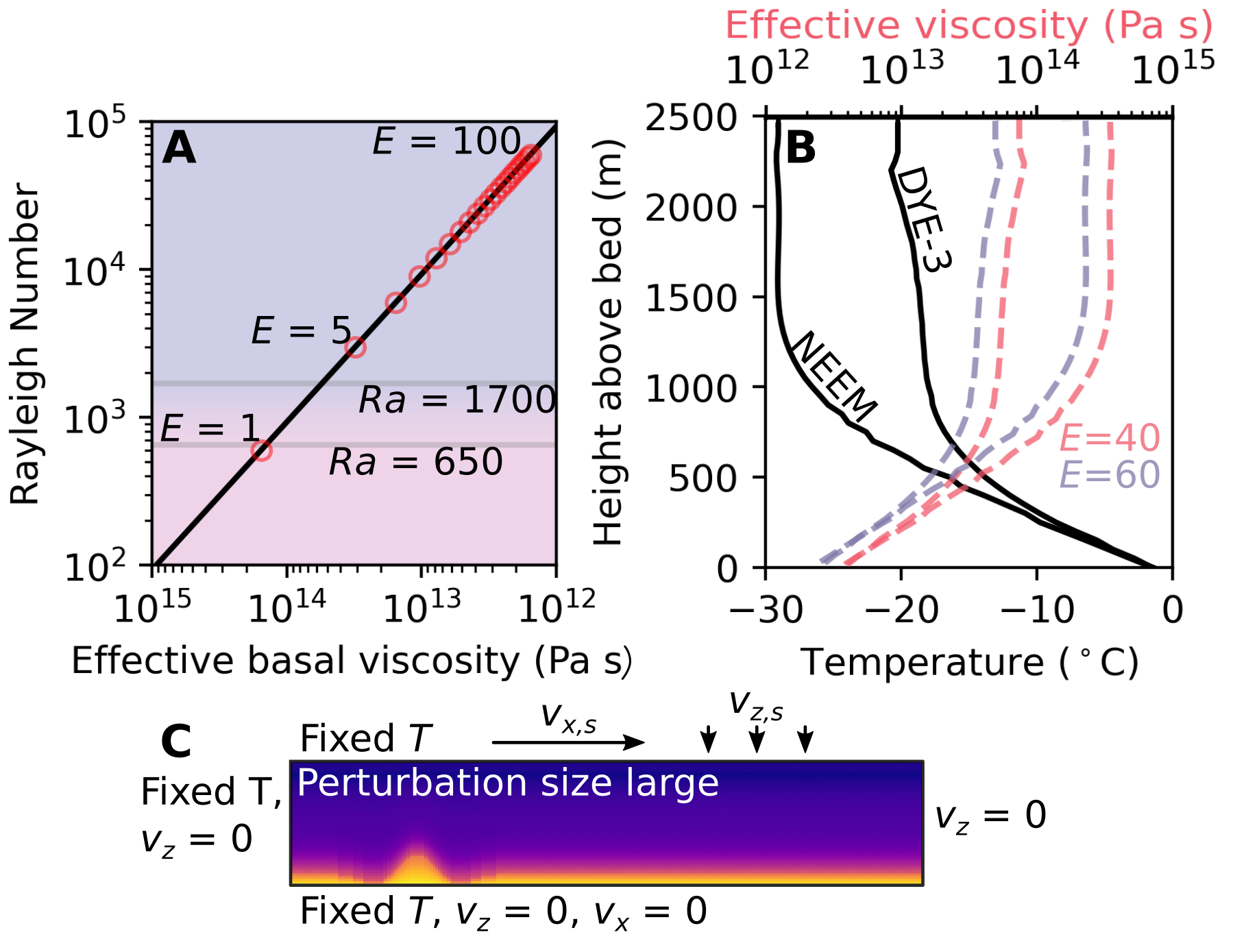}
\caption{Parameter and model information. \textbf{A} Rayleigh number calculated assuming the basal effective viscosity is constant through an ice column of 2,500 m thickness and how enhancement factors, \(E\), relate to a given effective basal viscosity when \(\tau_e=5\times10^4\) Pa. The lower and upper gray lines show Rayleigh numbers of 650 and 1700, respectively. \textbf{B} Temperature profiles from DYE-3 and NEEM \cite{Dahl-Jensen2013EemianCore, Rasmussen2013ACore} and attendant effective viscosity profiles given enhancement factors of 40 (red) and 60 (blue). \textbf{C} Initial model domain showing boundary conditions with a vertical enhancement factor of 2, length of 25 km and height of 2,500 m and a large temperature perturbation for the NEEM profile. The plasma colormap corresponds to the temperature colorbar in Figs. 1, 4. The medium temperature perturbation and 3D setup are shown in Fig. S4. }
\label{fig:misc}
\end{figure*}

\begin{figure*}
\noindent\includegraphics[width=0.5\textwidth]{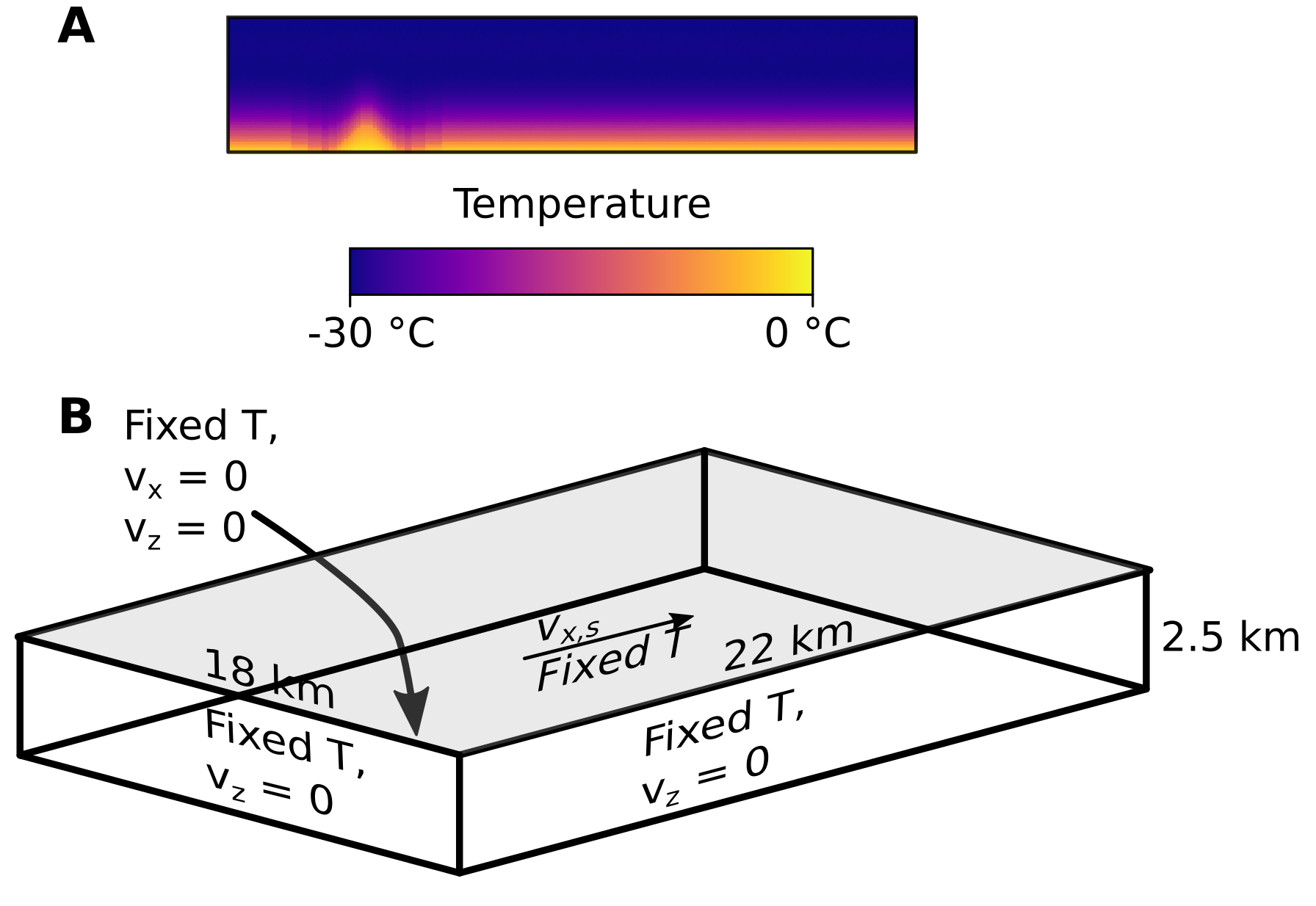}
\caption{A. The medium temperature perturbation, at the same scale as Fig. 3. B. Boundary setup for 3D simulations.}
\label{fig:misc_2}
\end{figure*}

\begin{figure}
\noindent\includegraphics[width=0.8\textwidth]{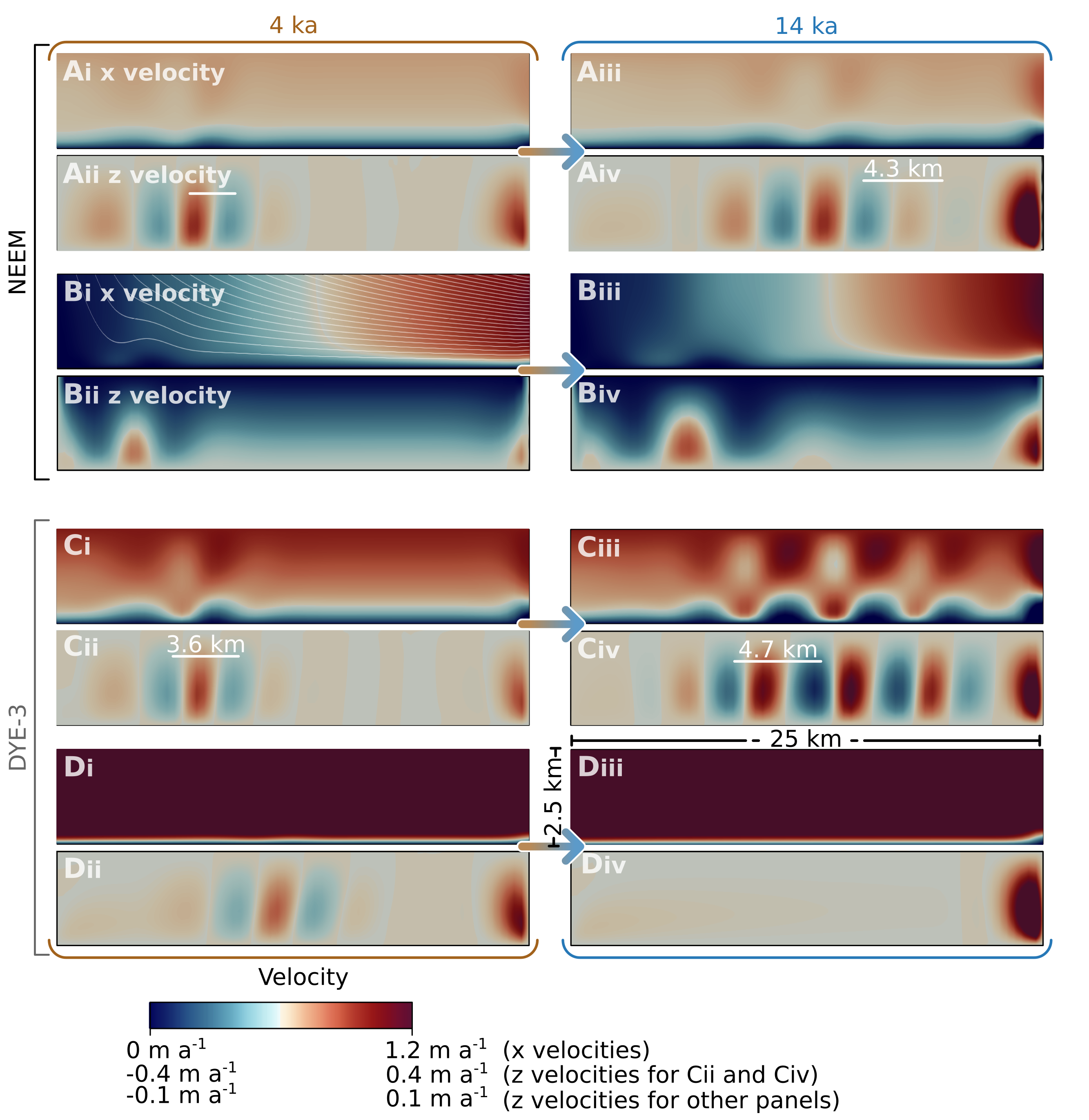}
\caption{As for Fig. 5 but showing x and z velocity components. Panel Bi additionally shows flow lines originating from the surface at a spacing of 625 m.}
\label{fig:plumes_out_2}
\end{figure}

\begin{figure*}
\noindent\includegraphics[width=0.3\textwidth]{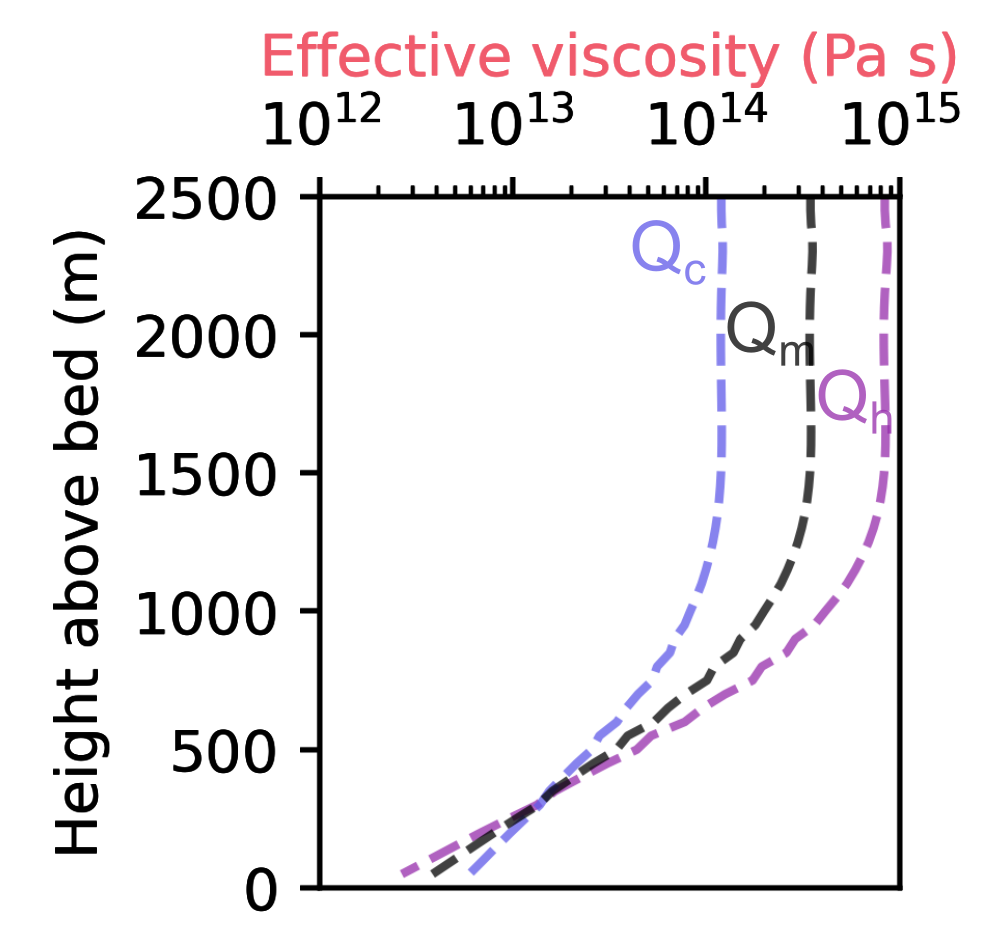}
\caption{Effect of different \(Q\) values on effective viscosity for \(E=40\) and the NEEM temperature profile. \(Q_c=6\times10^4\) is for \(T<-10\)\degree C, \(Q_h=11.5\times10^4\) is for \(T\geq-10\)\degree C, and \(Q_m=9\times10^4\) is the midpoint used in this study.}
\label{fig:Q_comp}
\end{figure*}

\begin{figure*}
\noindent\includegraphics[width=0.4\textwidth]{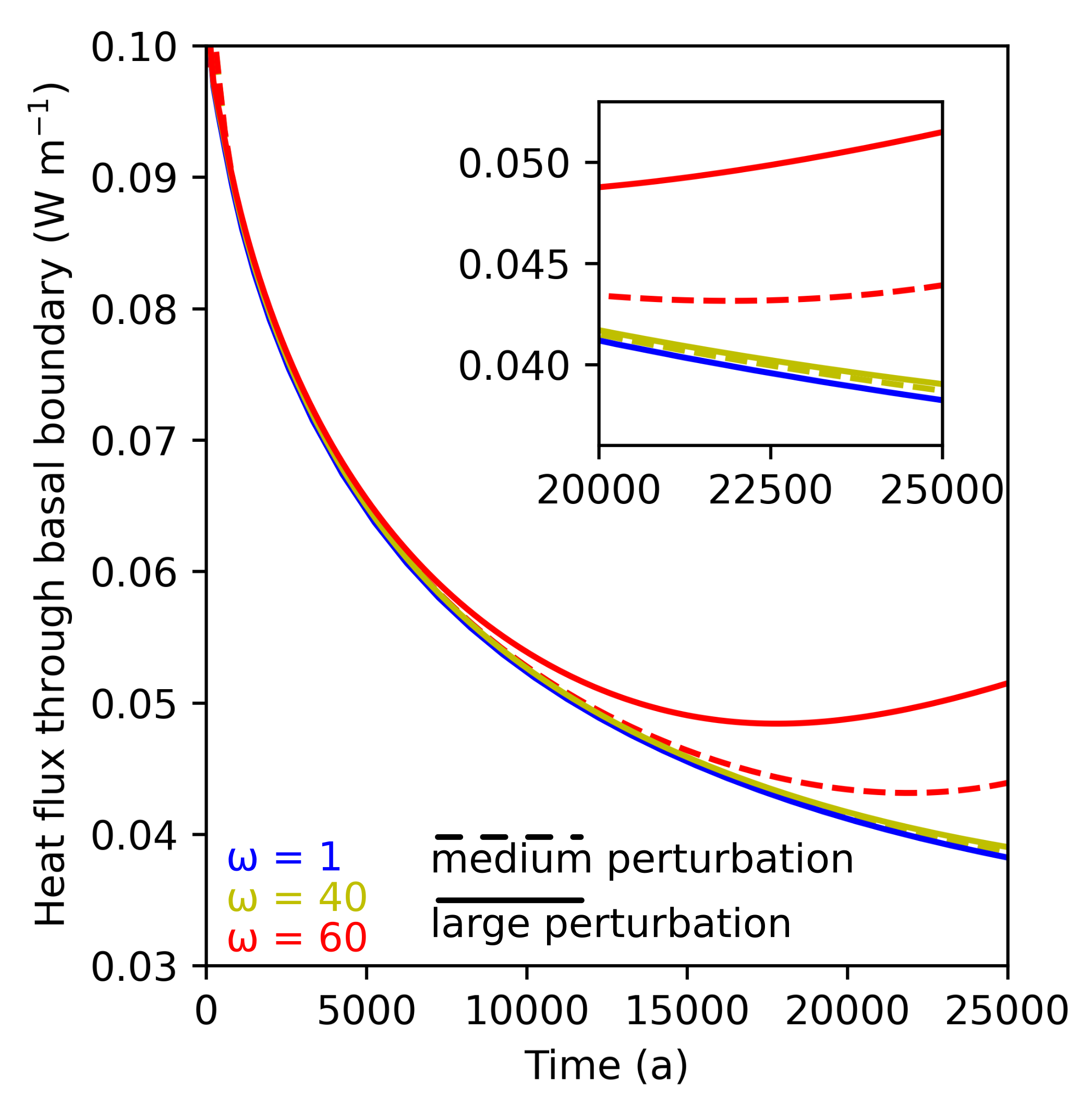}
\caption{Heat flux through the basal boundary for simulations run with the NEEM temperature profile, an ice thickness of 2,500 m, and 0 surface velocity. The dashed line refers to the medium perturbation, while the solid line refers to the large perturbation. A detail panel with the same axis is shown as an inset. The initial decrease in heat flux in all runs is a result of the basal temperature gradient (Fig. 3) reducing over time, before convection increases the heat flux again in simulations where the enhancement factor is 40 and 60. The medium perturbation results in a lower heat flux at a given time after \(\sim\)10 ka as the large perturbation essentially gives the convection state a head start}
\label{fig:ghf}
\end{figure*}

\begin{table}
\caption{Set parameters\label{table:parameters}}
\centering
\begin{tabular}{l c}
\hline
 Parameter  & Value  \\
\hline
  Activation energy, $Q$  & \(9\times10^4\) (kJ mol\textsuperscript{-1})   \\
  Ideal gas constant, $R$  & 8.314 J K$^{-1}$ mol$^{-1}$  \\
  Creep prefactor, $A_0$  & \(3.5\times 10^{-25}\) Pa s   \\
  Flow exponent, \(n\) & 3 \\
  Specific heat capacity, $C_p$  & 270 J\,kg$^{-1}$\,K$^{-1}$ \\
  Reference density, $\bar{\rho}$  & 917 kg m\textsuperscript{-3}   \\
  Thermal expansion coefficient, $\beta$  & 380 K\textsuperscript{-1}  \\
  Thermal conductivity, $\kappa$  & 370  W\,m$^{-1}$\,K$^{-1}$  \\
  Thermal diffusivity, \(\alpha\) & \(3.5\times 10^{-6}\) m\textsuperscript{2} s\textsuperscript{-1} \\
\hline
%\multicolumn{2}{l}{$^{a}$Footnote text here.}
\end{tabular}
\end{table}

\begin{table}
\caption{Run setups corresponding to panels in Fig. 4. Format of e.g. 25:5:100 indicates steps from 25 to 100 in spacing increments of 5. Run format of e.g. B-3Di corresponds to the first 3D run within the parameter space of B. B-3Div is equivalent to F-3Dii. \label{table:runs}}
\begin{tabular}{ccccccc}
Run & $E$ & $v_{x,s}$ & $H$ & $v_{z,s}$ & T profile & Perturbation \\
\hline
A & 25:5:100 & 1 & 2,500 & 0 & DYE-3 & large \\ 
B & 25:5:100 & 1 & 2,500 & 0 & NEEM & large \\
C & 25:5:100 & 1 & 2,500 & 0 & NEEM & medium \\
D & 40 & 0:0.2:3 & 2,500 & 0 & DYE-3 & large \\
E & 40 & 0:0.2:3 & 2,500 & 0 & NEEM & large \\
F & 60 & 0:0.2:3 & 2,500 & 0 & NEEM & large \\
G & 40 & 1 & 1,800:100:3,500 & 0 & DYE-3 & large \\
H & 40 & 1 & 1,800:100:3,500 & 0 & NEEM & large \\
I & 60 & 1 & 1,800:100:3,500 & 0 & NEEM & large \\
J & 40 & N/A & 2,500 & $\begin{array}{c} \text{0.02:0.02:0.12,} \\ \text{0.15:0.05:0.4} \end{array}$ & DYE-3 & large \\
K & 40 & N/A & 2,500 & $\begin{array}{c} \text{0.02:0.02:0.12,} \\ \text{0.15:0.05:0.4} \end{array}$ & NEEM & large \\
L & 60 & N/A & 2,500 & $\begin{array}{c} \text{0.02:0.02:0.12,} \\ \text{0.15:0.05:0.4} \end{array}$ & NEEM & large \\
B-3Di & 30 & 1 & 2,500 & 0 & NEEM & 3D \\
B-3Dii & 40 & 1 & 2,500 & 0 & NEEM & 3D \\
B-3Diii & 50 & 1 & 2,500 & 0 & NEEM & 3D \\
B-3Div = F-3Dii & 60 & 1 & 2,500 & 0 & NEEM & 3D \\
F-3Di & 60 & 0.7 & 2,500 & 0 & NEEM & 3D \\
F-3Diii & 60 & 1.3 & 2,500 & 0 & NEEM & 3D \\
\end{tabular}
\end{table}

\bibliography{references}

\begin{thebibliography}{}

\bibitem [\protect \citeauthoryear {%
Aschwanden%
\ \BBA {} Brinkerhoff%
}{%
Aschwanden%
\ \BBA {} Brinkerhoff%
}{%
{\protect \APACyear {2022}}%
}]{%
Aschwanden2022CalibratedSheet}
\APACinsertmetastar {%
Aschwanden2022CalibratedSheet}%
\begin{APACrefauthors}%
Aschwanden, A.%
\BCBT {}\ \BBA {} Brinkerhoff, D\BPBI J.%
\end{APACrefauthors}%
\unskip\
\newblock
\APACrefYearMonthDay{2022}{10}{}.
\newblock
{\BBOQ}\APACrefatitle {{Calibrated Mass Loss Predictions for the Greenland Ice Sheet}} {{Calibrated Mass Loss Predictions for the Greenland Ice Sheet}}.{\BBCQ}
\newblock
\APACjournalVolNumPages{Geophysical Research Letters}{49}{19}{e2022GL099058}.
\newblock
\begin{APACrefDOI} \doi{10.1029/2022GL099058} \end{APACrefDOI}
\PrintBackRefs{\CurrentBib}

\bibitem [\protect \citeauthoryear {%
Bangerth%
\ \protect \BOthers {.}}{%
Bangerth%
\ \protect \BOthers {.}}{%
{\protect \APACyear {2023}}%
}]{%
Bangerth2023ASPECT2.5.0}
\APACinsertmetastar {%
Bangerth2023ASPECT2.5.0}%
\begin{APACrefauthors}%
Bangerth, W.%
, Dannberg, J.%
, Fraters, M.%
, Gassmoeller, R.%
, Glerum, A.%
, Heister, T.%
\BDBL {}Naliboff, J.%
\end{APACrefauthors}%
\unskip\
\newblock
\APACrefYearMonthDay{2023}{}{}.
\newblock
\APACrefbtitle {{ASPECT 2.5.0}.} {{ASPECT 2.5.0}.}
\newblock
\APACaddressPublisher{}{Zenodo}.
\PrintBackRefs{\CurrentBib}

\bibitem [\protect \citeauthoryear {%
Bell%
\ \protect \BOthers {.}}{%
Bell%
\ \protect \BOthers {.}}{%
{\protect \APACyear {2011}}%
}]{%
Bell2011WidespreadBase}
\APACinsertmetastar {%
Bell2011WidespreadBase}%
\begin{APACrefauthors}%
Bell, R\BPBI E.%
, Ferraccioli, F.%
, Creyts, T\BPBI T.%
, Braaten, D.%
, Corr, H.%
, Das, I.%
\BDBL {}Wolovick, M.%
\end{APACrefauthors}%
\unskip\
\newblock
\APACrefYearMonthDay{2011}{3}{}.
\newblock
{\BBOQ}\APACrefatitle {{Widespread Persistent Thickening of the East Antarctic Ice Sheet by Freezing from the Base}} {{Widespread Persistent Thickening of the East Antarctic Ice Sheet by Freezing from the Base}}.{\BBCQ}
\newblock
\APACjournalVolNumPages{Science}{331}{6024}{1592--1595}.
\newblock
\begin{APACrefDOI} \doi{10.1126/SCIENCE.1200109} \end{APACrefDOI}
\PrintBackRefs{\CurrentBib}

\bibitem [\protect \citeauthoryear {%
Bell%
\ \protect \BOthers {.}}{%
Bell%
\ \protect \BOthers {.}}{%
{\protect \APACyear {2014}}%
}]{%
Bell2014DeformationMeltwater}
\APACinsertmetastar {%
Bell2014DeformationMeltwater}%
\begin{APACrefauthors}%
Bell, R\BPBI E.%
, Tinto, K.%
, Das, I.%
, Wolovick, M.%
, Chu, W.%
, Creyts, T\BPBI T.%
\BDBL {}Paden, J\BPBI D.%
\end{APACrefauthors}%
\unskip\
\newblock
\APACrefYearMonthDay{2014}{7}{}.
\newblock
{\BBOQ}\APACrefatitle {{Deformation, warming and softening of Greenland’s ice by refreezing meltwater}} {{Deformation, warming and softening of Greenland’s ice by refreezing meltwater}}.{\BBCQ}
\newblock
\APACjournalVolNumPages{Nature Geoscience}{7}{7}{497--502}.
\newblock
\begin{APACrefURL} \url{http://www.nature.com/articles/ngeo2179} \end{APACrefURL}
\newblock
\begin{APACrefDOI} \doi{10.1038/ngeo2179} \end{APACrefDOI}
\PrintBackRefs{\CurrentBib}

\bibitem [\protect \citeauthoryear {%
Berends%
, van~de Wal%
, van~den Akker%
\BCBL {}\ \BBA {} Lipscomb%
}{%
Berends%
\ \protect \BOthers {.}}{%
{\protect \APACyear {2023}}%
}]{%
Berends2023CompensatingResponse}
\APACinsertmetastar {%
Berends2023CompensatingResponse}%
\begin{APACrefauthors}%
Berends, C\BPBI J.%
, van~de Wal, R\BPBI S\BPBI W.%
, van~den Akker, T.%
\BCBL {}\ \BBA {} Lipscomb, W\BPBI H.%
\end{APACrefauthors}%
\unskip\
\newblock
\APACrefYearMonthDay{2023}{4}{}.
\newblock
{\BBOQ}\APACrefatitle {{Compensating errors in inversions for subglacial bed roughness: same steady state, different dynamic response}} {{Compensating errors in inversions for subglacial bed roughness: same steady state, different dynamic response}}.{\BBCQ}
\newblock
\APACjournalVolNumPages{The Cryosphere}{17}{4}{1585--1600}.
\newblock
\begin{APACrefURL} \url{https://tc.copernicus.org/articles/17/1585/2023/} \end{APACrefURL}
\newblock
\begin{APACrefDOI} \doi{10.5194/TC-17-1585-2023} \end{APACrefDOI}
\PrintBackRefs{\CurrentBib}

\bibitem [\protect \citeauthoryear {%
Bingham%
\ \protect \BOthers {.}}{%
Bingham%
\ \protect \BOthers {.}}{%
{\protect \APACyear {2024}}%
}]{%
Bingham2024ReviewSheets}
\APACinsertmetastar {%
Bingham2024ReviewSheets}%
\begin{APACrefauthors}%
Bingham, R\BPBI G.%
, Bodart, J\BPBI A.%
, Cavitte, M\BPBI G\BPBI P.%
, Chung, A.%
, Sanderson, R\BPBI J.%
, Sutter, J\BPBI C\BPBI R.%
\BDBL {}Zuhr, A.%
\end{APACrefauthors}%
\unskip\
\newblock
\APACrefYearMonthDay{2024}{10}{}.
\newblock
{\BBOQ}\APACrefatitle {{Review Article: Antarctica{\&}rsquo;s internal architecture: Towards a radiostratigraphically-informed age{\&}ndash;depth model of the Antarctic ice sheets}} {{Review Article: Antarctica{\&}rsquo;s internal architecture: Towards a radiostratigraphically-informed age{\&}ndash;depth model of the Antarctic ice sheets}}.{\BBCQ}
\newblock
\APACjournalVolNumPages{EGUsphere}{17}{}{1--66}.
\newblock
\begin{APACrefURL} \url{https://egusphere.copernicus.org/preprints/2024/egusphere-2024-2593/} \end{APACrefURL}
\newblock
\begin{APACrefDOI} \doi{10.5194/EGUSPHERE-2024-2593} \end{APACrefDOI}
\PrintBackRefs{\CurrentBib}

\bibitem [\protect \citeauthoryear {%
Bons%
\ \protect \BOthers {.}}{%
Bons%
\ \protect \BOthers {.}}{%
{\protect \APACyear {2016}}%
}]{%
Bons2016ConvergingSheet}
\APACinsertmetastar {%
Bons2016ConvergingSheet}%
\begin{APACrefauthors}%
Bons, P\BPBI D.%
, Jansen, D.%
, Mundel, F.%
, Bauer, C\BPBI C.%
, Binder, T.%
, Eisen, O.%
\BDBL {}Weikusat, I.%
\end{APACrefauthors}%
\unskip\
\newblock
\APACrefYearMonthDay{2016}{12}{}.
\newblock
{\BBOQ}\APACrefatitle {{Converging flow and anisotropy cause large-scale folding in Greenland's ice sheet}} {{Converging flow and anisotropy cause large-scale folding in Greenland's ice sheet}}.{\BBCQ}
\newblock
\APACjournalVolNumPages{Nature Communications}{7}{}{11427}.
\newblock
\begin{APACrefDOI} \doi{10.1038/ncomms11427} \end{APACrefDOI}
\PrintBackRefs{\CurrentBib}

\bibitem [\protect \citeauthoryear {%
Bons%
\ \protect \BOthers {.}}{%
Bons%
\ \protect \BOthers {.}}{%
{\protect \APACyear {2018}}%
}]{%
Bons2018GreenlandMotion}
\APACinsertmetastar {%
Bons2018GreenlandMotion}%
\begin{APACrefauthors}%
Bons, P\BPBI D.%
, Kleiner, T.%
, Llorens, M\BHBI G.%
, Prior, D\BPBI J.%
, Sachau, T.%
, Weikusat, I.%
\BCBL {}\ \BBA {} Jansen, D.%
\end{APACrefauthors}%
\unskip\
\newblock
\APACrefYearMonthDay{2018}{7}{}.
\newblock
{\BBOQ}\APACrefatitle {{Greenland Ice Sheet: Higher Nonlinearity of Ice Flow Significantly Reduces Estimated Basal Motion}} {{Greenland Ice Sheet: Higher Nonlinearity of Ice Flow Significantly Reduces Estimated Basal Motion}}.{\BBCQ}
\newblock
\APACjournalVolNumPages{Geophysical Research Letters}{45}{13}{6542--6548}.
\newblock
\begin{APACrefDOI} \doi{10.1029/2018GL078356} \end{APACrefDOI}
\PrintBackRefs{\CurrentBib}

\bibitem [\protect \citeauthoryear {%
Cavitte%
\ \protect \BOthers {.}}{%
Cavitte%
\ \protect \BOthers {.}}{%
{\protect \APACyear {2021}}%
}]{%
Cavitte2021AMid-Pleistocene}
\APACinsertmetastar {%
Cavitte2021AMid-Pleistocene}%
\begin{APACrefauthors}%
Cavitte, M\BPBI G.%
, Young, D\BPBI A.%
, Mulvaney, R.%
, Ritz, C.%
, Greenbaum, J\BPBI S.%
, Ng, G.%
\BDBL {}Blankenship, D\BPBI D.%
\end{APACrefauthors}%
\unskip\
\newblock
\APACrefYearMonthDay{2021}{10}{}.
\newblock
{\BBOQ}\APACrefatitle {{A detailed radiostratigraphic data set for the central East Antarctic Plateau spanning from the Holocene to the mid-Pleistocene}} {{A detailed radiostratigraphic data set for the central East Antarctic Plateau spanning from the Holocene to the mid-Pleistocene}}.{\BBCQ}
\newblock
\APACjournalVolNumPages{Earth System Science Data}{13}{10}{4759--4777}.
\newblock
\begin{APACrefDOI} \doi{10.5194/ESSD-13-4759-2021} \end{APACrefDOI}
\PrintBackRefs{\CurrentBib}

\bibitem [\protect \citeauthoryear {%
{CReSIS}%
}{%
{CReSIS}%
}{%
{\protect \APACyear {2013}}%
}]{%
CReSIS2013RadiostratigraphyCenter}
\APACinsertmetastar {%
CReSIS2013RadiostratigraphyCenter}%
\begin{APACrefauthors}%
{CReSIS}.%
\end{APACrefauthors}%
\unskip\
\newblock
\APACrefYearMonthDay{2013}{}{}.
\newblock
\APACrefbtitle {{Radiostratigraphy and Age Structure of the Greenland Ice Sheet, Version 1 | National Snow and Ice Data Center}.} {{Radiostratigraphy and Age Structure of the Greenland Ice Sheet, Version 1 | National Snow and Ice Data Center}.}
\newblock
\APACaddressPublisher{Colorado}{National Snow and Ice Data Center}.
\newblock
\begin{APACrefURL} \url{https://nsidc.org/data/rrrag4/versions/1} \end{APACrefURL}
\PrintBackRefs{\CurrentBib}

\bibitem [\protect \citeauthoryear {%
Cuffey%
\ \BBA {} Paterson%
}{%
Cuffey%
\ \BBA {} Paterson%
}{%
{\protect \APACyear {2010}}%
}]{%
Cuffey2010TheGlaciers}
\APACinsertmetastar {%
Cuffey2010TheGlaciers}%
\begin{APACrefauthors}%
Cuffey, K\BPBI M.%
\BCBT {}\ \BBA {} Paterson, W\BPBI S\BPBI B.%
\end{APACrefauthors}%
\unskip\
\newblock
\APACrefYear{2010}.
\newblock
\APACrefbtitle {{The Physics of Glaciers}} {{The Physics of Glaciers}}\ (\PrintOrdinal{4}\ \BEd).
\newblock
\APACaddressPublisher{Amsterdam}{Elsevier Science {\&} Technology Books}.
\PrintBackRefs{\CurrentBib}

\bibitem [\protect \citeauthoryear {%
Dahl-Jensen%
\ \protect \BOthers {.}}{%
Dahl-Jensen%
\ \protect \BOthers {.}}{%
{\protect \APACyear {2013}}%
}]{%
Dahl-Jensen2013EemianCore}
\APACinsertmetastar {%
Dahl-Jensen2013EemianCore}%
\begin{APACrefauthors}%
Dahl-Jensen, D.%
, Albert, M\BPBI R.%
, Aldahan, A.%
, Azuma, N.%
, Balslev-Clausen, D.%
, Baumgartner, M.%
\BDBL {}Zheng, J.%
\end{APACrefauthors}%
\unskip\
\newblock
\APACrefYearMonthDay{2013}{1}{}.
\newblock
{\BBOQ}\APACrefatitle {{Eemian interglacial reconstructed from a Greenland folded ice core}} {{Eemian interglacial reconstructed from a Greenland folded ice core}}.{\BBCQ}
\newblock
\APACjournalVolNumPages{Nature 2013 493:7433}{493}{7433}{489--494}.
\newblock
\begin{APACrefDOI} \doi{10.1038/nature11789} \end{APACrefDOI}
\PrintBackRefs{\CurrentBib}

\bibitem [\protect \citeauthoryear {%
Dow%
, Karlsson%
\BCBL {}\ \BBA {} Werder%
}{%
Dow%
\ \protect \BOthers {.}}{%
{\protect \APACyear {2018}}%
}]{%
Dow2018LimitedStratigraphy}
\APACinsertmetastar {%
Dow2018LimitedStratigraphy}%
\begin{APACrefauthors}%
Dow, C\BPBI F.%
, Karlsson, N\BPBI B.%
\BCBL {}\ \BBA {} Werder, M\BPBI A.%
\end{APACrefauthors}%
\unskip\
\newblock
\APACrefYearMonthDay{2018}{2}{}.
\newblock
{\BBOQ}\APACrefatitle {{Limited Impact of Subglacial Supercooling Freeze-on for Greenland Ice Sheet Stratigraphy}} {{Limited Impact of Subglacial Supercooling Freeze-on for Greenland Ice Sheet Stratigraphy}}.{\BBCQ}
\newblock
\APACjournalVolNumPages{Geophysical Research Letters}{45}{3}{1481--1489}.
\newblock
\begin{APACrefDOI} \doi{10.1002/2017GL076251} \end{APACrefDOI}
\PrintBackRefs{\CurrentBib}

\bibitem [\protect \citeauthoryear {%
Echelmeyer%
, Harrison%
, Larsen%
\BCBL {}\ \BBA {} Mitchell%
}{%
Echelmeyer%
\ \protect \BOthers {.}}{%
{\protect \APACyear {1994}}%
}]{%
Echelmeyer1994TheStream}
\APACinsertmetastar {%
Echelmeyer1994TheStream}%
\begin{APACrefauthors}%
Echelmeyer, K.%
, Harrison, W.%
, Larsen, C.%
\BCBL {}\ \BBA {} Mitchell, J.%
\end{APACrefauthors}%
\unskip\
\newblock
\APACrefYearMonthDay{1994}{}{}.
\newblock
{\BBOQ}\APACrefatitle {{The role of the margins in the dynamics of an active ice stream}} {{The role of the margins in the dynamics of an active ice stream}}.{\BBCQ}
\newblock
\APACjournalVolNumPages{Journal of Glaciology}{40}{136}{527--538}.
\newblock
\begin{APACrefDOI} \doi{10.3189/S0022143000012417} \end{APACrefDOI}
\PrintBackRefs{\CurrentBib}

\bibitem [\protect \citeauthoryear {%
Echelmeyer%
\ \BBA {} Zhongxiang%
}{%
Echelmeyer%
\ \BBA {} Zhongxiang%
}{%
{\protect \APACyear {1987}}%
}]{%
Echelmeyer1987DirectTemperatures}
\APACinsertmetastar {%
Echelmeyer1987DirectTemperatures}%
\begin{APACrefauthors}%
Echelmeyer, K.%
\BCBT {}\ \BBA {} Zhongxiang, W.%
\end{APACrefauthors}%
\unskip\
\newblock
\APACrefYearMonthDay{1987}{}{}.
\newblock
{\BBOQ}\APACrefatitle {{Direct Observation of Basal Sliding and Deformation of Basal Drift at Sub-Freezing Temperatures}} {{Direct Observation of Basal Sliding and Deformation of Basal Drift at Sub-Freezing Temperatures}}.{\BBCQ}
\newblock
\APACjournalVolNumPages{Journal of Glaciology}{33}{113}{83--98}.
\newblock
\begin{APACrefDOI} \doi{10.3189/S0022143000005396} \end{APACrefDOI}
\PrintBackRefs{\CurrentBib}

\bibitem [\protect \citeauthoryear {%
Fortuin%
\ \BBA {} Oerlemans%
}{%
Fortuin%
\ \BBA {} Oerlemans%
}{%
{\protect \APACyear {1990}}%
}]{%
Fortuin1990ParameterizationAntarctica}
\APACinsertmetastar {%
Fortuin1990ParameterizationAntarctica}%
\begin{APACrefauthors}%
Fortuin, J\BPBI P.%
\BCBT {}\ \BBA {} Oerlemans, J.%
\end{APACrefauthors}%
\unskip\
\newblock
\APACrefYearMonthDay{1990}{}{}.
\newblock
{\BBOQ}\APACrefatitle {{Parameterization of the Annual Surface Temperature and Mass Balance of Antarctica}} {{Parameterization of the Annual Surface Temperature and Mass Balance of Antarctica}}.{\BBCQ}
\newblock
\APACjournalVolNumPages{Annals of Glaciology}{14}{}{78--84}.
\newblock
\begin{APACrefDOI} \doi{10.3189/S0260305500008302} \end{APACrefDOI}
\PrintBackRefs{\CurrentBib}

\bibitem [\protect \citeauthoryear {%
Fowler%
}{%
Fowler%
}{%
{\protect \APACyear {2013}}%
}]{%
Fowler2013ThermalSheets}
\APACinsertmetastar {%
Fowler2013ThermalSheets}%
\begin{APACrefauthors}%
Fowler, A\BPBI C.%
\end{APACrefauthors}%
\unskip\
\newblock
\APACrefYearMonthDay{2013}{3}{}.
\newblock
{\BBOQ}\APACrefatitle {{Thermal convection in ice sheets}} {{Thermal convection in ice sheets}}.{\BBCQ}
\newblock
\APACjournalVolNumPages{Journal of Glaciology}{59}{213}{190--192}.
\newblock
\begin{APACrefDOI} \doi{10.3189/2013JOG12J164} \end{APACrefDOI}
\PrintBackRefs{\CurrentBib}

\bibitem [\protect \citeauthoryear {%
Glen%
}{%
Glen%
}{%
{\protect \APACyear {1955}}%
}]{%
Glen1955TheIce}
\APACinsertmetastar {%
Glen1955TheIce}%
\begin{APACrefauthors}%
Glen, J\BPBI W.%
\end{APACrefauthors}%
\unskip\
\newblock
\APACrefYearMonthDay{1955}{3}{}.
\newblock
{\BBOQ}\APACrefatitle {{The creep of polycrystalline ice}} {{The creep of polycrystalline ice}}.{\BBCQ}
\newblock
\APACjournalVolNumPages{Proceedings of the Royal Society of London. Series A. Mathematical and Physical Sciences}{228}{1175}{519--538}.
\newblock
\begin{APACrefDOI} \doi{10.1098/rspa.1955.0066} \end{APACrefDOI}
\PrintBackRefs{\CurrentBib}

\bibitem [\protect \citeauthoryear {%
Goelzer%
\ \protect \BOthers {.}}{%
Goelzer%
\ \protect \BOthers {.}}{%
{\protect \APACyear {2020}}%
}]{%
Goelzer2020TheISMIP6}
\APACinsertmetastar {%
Goelzer2020TheISMIP6}%
\begin{APACrefauthors}%
Goelzer, H.%
, Nowicki, S.%
, Payne, A.%
, Larour, E.%
, Seroussi, H.%
, Lipscomb, W\BPBI H.%
\BDBL {}Aschwanden, A.%
\end{APACrefauthors}%
\unskip\
\newblock
\APACrefYearMonthDay{2020}{9}{}.
\newblock
{\BBOQ}\APACrefatitle {{The future sea-level contribution of the Greenland ice sheet: A multi-model ensemble study of ISMIP6}} {{The future sea-level contribution of the Greenland ice sheet: A multi-model ensemble study of ISMIP6}}.{\BBCQ}
\newblock
\APACjournalVolNumPages{The Cryosphere}{14}{9}{3071--3096}.
\newblock
\begin{APACrefDOI} \doi{10.5194/TC-14-3071-2020} \end{APACrefDOI}
\PrintBackRefs{\CurrentBib}

\bibitem [\protect \citeauthoryear {%
Gundestrup%
\ \BBA {} Hansen%
}{%
Gundestrup%
\ \BBA {} Hansen%
}{%
{\protect \APACyear {1984}}%
}]{%
Gundestrup1984Bore-HoleGreenland}
\APACinsertmetastar {%
Gundestrup1984Bore-HoleGreenland}%
\begin{APACrefauthors}%
Gundestrup, N\BPBI S.%
\BCBT {}\ \BBA {} Hansen, B\BPBI L.%
\end{APACrefauthors}%
\unskip\
\newblock
\APACrefYearMonthDay{1984}{}{}.
\newblock
{\BBOQ}\APACrefatitle {{Bore-Hole Survey at Dye 3, South Greenland}} {{Bore-Hole Survey at Dye 3, South Greenland}}.{\BBCQ}
\newblock
\APACjournalVolNumPages{Journal of Glaciology}{30}{106}{282--288}.
\newblock
\begin{APACrefDOI} \doi{10.3189/S0022143000006109} \end{APACrefDOI}
\PrintBackRefs{\CurrentBib}

\bibitem [\protect \citeauthoryear {%
Heister%
, Dannberg%
, Gassm{\"{o}}ller%
\BCBL {}\ \BBA {} Bangerth%
}{%
Heister%
\ \protect \BOthers {.}}{%
{\protect \APACyear {2017}}%
}]{%
Heister2017HighProblems}
\APACinsertmetastar {%
Heister2017HighProblems}%
\begin{APACrefauthors}%
Heister, T.%
, Dannberg, J.%
, Gassm{\"{o}}ller, R.%
\BCBL {}\ \BBA {} Bangerth, W.%
\end{APACrefauthors}%
\unskip\
\newblock
\APACrefYearMonthDay{2017}{8}{}.
\newblock
{\BBOQ}\APACrefatitle {{High accuracy mantle convection simulation through modern numerical methods – II: realistic models and problems}} {{High accuracy mantle convection simulation through modern numerical methods – II: realistic models and problems}}.{\BBCQ}
\newblock
\APACjournalVolNumPages{Geophysical Journal International}{210}{2}{833--851}.
\newblock
\begin{APACrefDOI} \doi{10.1093/GJI/GGX195} \end{APACrefDOI}
\PrintBackRefs{\CurrentBib}

\bibitem [\protect \citeauthoryear {%
Hughes%
}{%
Hughes%
}{%
{\protect \APACyear {1976}}%
}]{%
Hughes1976TheSheets}
\APACinsertmetastar {%
Hughes1976TheSheets}%
\begin{APACrefauthors}%
Hughes, T.%
\end{APACrefauthors}%
\unskip\
\newblock
\APACrefYearMonthDay{1976}{}{}.
\newblock
{\BBOQ}\APACrefatitle {{The Theory of Thermal Convection in Polar Ice Sheets}} {{The Theory of Thermal Convection in Polar Ice Sheets}}.{\BBCQ}
\newblock
\APACjournalVolNumPages{Journal of Glaciology}{16}{74}{41--71}.
\newblock
\begin{APACrefDOI} \doi{10.3189/S0022143000031427} \end{APACrefDOI}
\PrintBackRefs{\CurrentBib}

\bibitem [\protect \citeauthoryear {%
Hughes%
}{%
Hughes%
}{%
{\protect \APACyear {2012}}%
}]{%
Hughes2012AreSheet}
\APACinsertmetastar {%
Hughes2012AreSheet}%
\begin{APACrefauthors}%
Hughes, T.%
\end{APACrefauthors}%
\unskip\
\newblock
\APACrefYearMonthDay{2012}{8}{}.
\newblock
{\BBOQ}\APACrefatitle {{Are ice-stream tributaries the surface expression of thermal convection rolls in the Antarctic ice sheet?}} {{Are ice-stream tributaries the surface expression of thermal convection rolls in the Antarctic ice sheet?}}{\BBCQ}
\newblock
\APACjournalVolNumPages{Journal of Glaciology}{58}{210}{811--814}.
\newblock
\begin{APACrefDOI} \doi{10.3189/2012JOG12J068} \end{APACrefDOI}
\PrintBackRefs{\CurrentBib}

\bibitem [\protect \citeauthoryear {%
Knopoff%
}{%
Knopoff%
}{%
{\protect \APACyear {1964}}%
}]{%
Knopoff1964TheHypothesis}
\APACinsertmetastar {%
Knopoff1964TheHypothesis}%
\begin{APACrefauthors}%
Knopoff, L.%
\end{APACrefauthors}%
\unskip\
\newblock
\APACrefYearMonthDay{1964}{2}{}.
\newblock
{\BBOQ}\APACrefatitle {{The convection current hypothesis}} {{The convection current hypothesis}}.{\BBCQ}
\newblock
\APACjournalVolNumPages{Reviews of Geophysics}{2}{1}{89--122}.
\newblock
\begin{APACrefDOI} \doi{10.1029/RG002I001P00089} \end{APACrefDOI}
\PrintBackRefs{\CurrentBib}

\bibitem [\protect \citeauthoryear {%
Kronbichler%
, Heister%
\BCBL {}\ \BBA {} Bangerth%
}{%
Kronbichler%
\ \protect \BOthers {.}}{%
{\protect \APACyear {2012}}%
}]{%
Kronbichler2012HighMethods}
\APACinsertmetastar {%
Kronbichler2012HighMethods}%
\begin{APACrefauthors}%
Kronbichler, M.%
, Heister, T.%
\BCBL {}\ \BBA {} Bangerth, W.%
\end{APACrefauthors}%
\unskip\
\newblock
\APACrefYearMonthDay{2012}{10}{}.
\newblock
{\BBOQ}\APACrefatitle {{High accuracy mantle convection simulation through modern numerical methods}} {{High accuracy mantle convection simulation through modern numerical methods}}.{\BBCQ}
\newblock
\APACjournalVolNumPages{Geophysical Journal International}{191}{1}{12--29}.
\newblock
\begin{APACrefURL} \url{https://onlinelibrary.wiley.com/doi/full/10.1111/j.1365-246X.2012.05609.x https://onlinelibrary.wiley.com/doi/abs/10.1111/j.1365-246X.2012.05609.x https://onlinelibrary.wiley.com/doi/10.1111/j.1365-246X.2012.05609.x} \end{APACrefURL}
\newblock
\begin{APACrefDOI} \doi{10.1111/J.1365-246X.2012.05609.X} \end{APACrefDOI}
\PrintBackRefs{\CurrentBib}

\bibitem [\protect \citeauthoryear {%
Larour%
, Seroussi%
, Morlighem%
\BCBL {}\ \BBA {} Rignot%
}{%
Larour%
\ \protect \BOthers {.}}{%
{\protect \APACyear {2012}}%
}]{%
Larour2012ContinentalISSM}
\APACinsertmetastar {%
Larour2012ContinentalISSM}%
\begin{APACrefauthors}%
Larour, E.%
, Seroussi, H.%
, Morlighem, M.%
\BCBL {}\ \BBA {} Rignot, E.%
\end{APACrefauthors}%
\unskip\
\newblock
\APACrefYearMonthDay{2012}{3}{}.
\newblock
{\BBOQ}\APACrefatitle {{Continental scale, high order, high spatial resolution, ice sheet modeling using the Ice Sheet System Model (ISSM)}} {{Continental scale, high order, high spatial resolution, ice sheet modeling using the Ice Sheet System Model (ISSM)}}.{\BBCQ}
\newblock
\APACjournalVolNumPages{Journal of Geophysical Research: Earth Surface}{117}{F1}{1022}.
\newblock
\begin{APACrefDOI} \doi{10.1029/2011JF002140} \end{APACrefDOI}
\PrintBackRefs{\CurrentBib}

\bibitem [\protect \citeauthoryear {%
Law%
\ \protect \BOthers {.}}{%
Law%
\ \protect \BOthers {.}}{%
{\protect \APACyear {2021}}%
}]{%
Law2021ThermodynamicsSensing}
\APACinsertmetastar {%
Law2021ThermodynamicsSensing}%
\begin{APACrefauthors}%
Law, R.%
, Christoffersen, P.%
, Hubbard, B.%
, Doyle, S\BPBI H.%
, Chudley, T\BPBI R.%
, Schoonman, C.%
\BDBL {}Young, T\BPBI J.%
\end{APACrefauthors}%
\unskip\
\newblock
\APACrefYearMonthDay{2021}{}{}.
\newblock
{\BBOQ}\APACrefatitle {{Thermodynamics of a fast-moving Greenlandic outlet glacier revealed by fiber-optic distributed temperature sensing}} {{Thermodynamics of a fast-moving Greenlandic outlet glacier revealed by fiber-optic distributed temperature sensing}}.{\BBCQ}
\newblock
\APACjournalVolNumPages{Science Advances}{7}{20}{eabe7136}.
\newblock
\begin{APACrefDOI} \doi{https://doi.org/10.1126/sciadv.abe7136} \end{APACrefDOI}
\PrintBackRefs{\CurrentBib}

\bibitem [\protect \citeauthoryear {%
Lebec%
, Labrosse%
, Morison%
\BCBL {}\ \BBA {} Tackley%
}{%
Lebec%
\ \protect \BOthers {.}}{%
{\protect \APACyear {2023}}%
}]{%
Lebec2023ScalingHabitability}
\APACinsertmetastar {%
Lebec2023ScalingHabitability}%
\begin{APACrefauthors}%
Lebec, L.%
, Labrosse, S.%
, Morison, A.%
\BCBL {}\ \BBA {} Tackley, P\BPBI J.%
\end{APACrefauthors}%
\unskip\
\newblock
\APACrefYearMonthDay{2023}{5}{}.
\newblock
{\BBOQ}\APACrefatitle {{Scaling of convection in high-pressure ice layers of large icy moons and implications for habitability}} {{Scaling of convection in high-pressure ice layers of large icy moons and implications for habitability}}.{\BBCQ}
\newblock
\APACjournalVolNumPages{Icarus}{396}{}{115494}.
\newblock
\begin{APACrefDOI} \doi{10.1016/J.ICARUS.2023.115494} \end{APACrefDOI}
\PrintBackRefs{\CurrentBib}

\bibitem [\protect \citeauthoryear {%
Lecavalier%
\ \protect \BOthers {.}}{%
Lecavalier%
\ \protect \BOthers {.}}{%
{\protect \APACyear {2014}}%
}]{%
Lecavalier2014AExtent}
\APACinsertmetastar {%
Lecavalier2014AExtent}%
\begin{APACrefauthors}%
Lecavalier, B\BPBI S.%
, Milne, G\BPBI A.%
, Simpson, M\BPBI J.%
, Wake, L.%
, Huybrechts, P.%
, Tarasov, L.%
\BDBL {}Larsen, N\BPBI K.%
\end{APACrefauthors}%
\unskip\
\newblock
\APACrefYearMonthDay{2014}{10}{}.
\newblock
{\BBOQ}\APACrefatitle {{A model of Greenland ice sheet deglaciation constrained by observations of relative sea level and ice extent}} {{A model of Greenland ice sheet deglaciation constrained by observations of relative sea level and ice extent}}.{\BBCQ}
\newblock
\APACjournalVolNumPages{Quaternary Science Reviews}{102}{}{54--84}.
\newblock
\begin{APACrefDOI} \doi{10.1016/J.QUASCIREV.2014.07.018} \end{APACrefDOI}
\PrintBackRefs{\CurrentBib}

\bibitem [\protect \citeauthoryear {%
Leysinger~Vieli%
, Mart{\'{i}}n%
, Hindmarsh%
\BCBL {}\ \BBA {} L{\"{u}}thi%
}{%
Leysinger~Vieli%
\ \protect \BOthers {.}}{%
{\protect \APACyear {2018}}%
}]{%
LeysingerVieli2018BasalStratigraphy}
\APACinsertmetastar {%
LeysingerVieli2018BasalStratigraphy}%
\begin{APACrefauthors}%
Leysinger~Vieli, G\BPBI J.%
, Mart{\'{i}}n, C.%
, Hindmarsh, R\BPBI C.%
\BCBL {}\ \BBA {} L{\"{u}}thi, M\BPBI P.%
\end{APACrefauthors}%
\unskip\
\newblock
\APACrefYearMonthDay{2018}{11}{}.
\newblock
{\BBOQ}\APACrefatitle {{Basal freeze-on generates complex ice-sheet stratigraphy}} {{Basal freeze-on generates complex ice-sheet stratigraphy}}.{\BBCQ}
\newblock
\APACjournalVolNumPages{Nature Communications 2018 9:1}{9}{1}{1--13}.
\newblock
\begin{APACrefDOI} \doi{10.1038/s41467-018-07083-3} \end{APACrefDOI}
\PrintBackRefs{\CurrentBib}

\bibitem [\protect \citeauthoryear {%
Macgregor%
\ \protect \BOthers {.}}{%
Macgregor%
\ \protect \BOthers {.}}{%
{\protect \APACyear {2022}}%
}]{%
Macgregor2022GBaTSv2:Sheet}
\APACinsertmetastar {%
Macgregor2022GBaTSv2:Sheet}%
\begin{APACrefauthors}%
Macgregor, J\BPBI A.%
, Chu, W.%
, Colgan, W\BPBI T.%
, Fahnestock, M\BPBI A.%
, Felikson, D.%
, Karlsson, N\BPBI B.%
\BDBL {}Studinger, M.%
\end{APACrefauthors}%
\unskip\
\newblock
\APACrefYearMonthDay{2022}{8}{}.
\newblock
{\BBOQ}\APACrefatitle {{GBaTSv2: a revised synthesis of the likely basal thermal state of the Greenland Ice Sheet}} {{GBaTSv2: a revised synthesis of the likely basal thermal state of the Greenland Ice Sheet}}.{\BBCQ}
\newblock
\APACjournalVolNumPages{Cryosphere}{16}{8}{3033--3049}.
\newblock
\begin{APACrefDOI} \doi{10.5194/TC-16-3033-2022} \end{APACrefDOI}
\PrintBackRefs{\CurrentBib}

\bibitem [\protect \citeauthoryear {%
MacGregor%
\ \protect \BOthers {.}}{%
MacGregor%
\ \protect \BOthers {.}}{%
{\protect \APACyear {2016}}%
}]{%
MacGregor2016HoloceneSheet}
\APACinsertmetastar {%
MacGregor2016HoloceneSheet}%
\begin{APACrefauthors}%
MacGregor, J\BPBI A.%
, Colgan, W\BPBI T.%
, Fahnestock, M\BPBI A.%
, Morlighem, M.%
, Catania, G\BPBI A.%
, Paden, J\BPBI D.%
\BCBL {}\ \BBA {} Gogineni, S\BPBI P.%
\end{APACrefauthors}%
\unskip\
\newblock
\APACrefYearMonthDay{2016}{2}{}.
\newblock
{\BBOQ}\APACrefatitle {{Holocene deceleration of the Greenland Ice Sheet}} {{Holocene deceleration of the Greenland Ice Sheet}}.{\BBCQ}
\newblock
\APACjournalVolNumPages{Science}{351}{6273}{590--593}.
\newblock
\begin{APACrefDOI} \doi{https://doi.org/10.1126/science.aab1702} \end{APACrefDOI}
\PrintBackRefs{\CurrentBib}

\bibitem [\protect \citeauthoryear {%
Macgregor%
\ \protect \BOthers {.}}{%
Macgregor%
\ \protect \BOthers {.}}{%
{\protect \APACyear {2015}}%
}]{%
Macgregor2015RadarSheet}
\APACinsertmetastar {%
Macgregor2015RadarSheet}%
\begin{APACrefauthors}%
Macgregor, J\BPBI A.%
, Li, J.%
, Paden, J\BPBI D.%
, Catania, G\BPBI A.%
, Clow, G\BPBI D.%
, Fahnestock, M\BPBI A.%
\BDBL {}Stillman, D\BPBI E.%
\end{APACrefauthors}%
\unskip\
\newblock
\APACrefYearMonthDay{2015}{6}{}.
\newblock
{\BBOQ}\APACrefatitle {{Radar attenuation and temperature within the Greenland Ice Sheet}} {{Radar attenuation and temperature within the Greenland Ice Sheet}}.{\BBCQ}
\newblock
\APACjournalVolNumPages{Journal of Geophysical Research: Earth Surface}{120}{6}{983--1008}.
\newblock
\begin{APACrefDOI} \doi{10.1002/2014JF003418} \end{APACrefDOI}
\PrintBackRefs{\CurrentBib}

\bibitem [\protect \citeauthoryear {%
Moon%
, Fisher%
, Stafford%
\BCBL {}\ \BBA {} Harden%
}{%
Moon%
\ \protect \BOthers {.}}{%
{\protect \APACyear {2022}}%
}]{%
Moon2022QGreenland:GIS}
\APACinsertmetastar {%
Moon2022QGreenland:GIS}%
\begin{APACrefauthors}%
Moon, T.%
, Fisher, M.%
, Stafford, T.%
\BCBL {}\ \BBA {} Harden, L.%
\end{APACrefauthors}%
\unskip\
\newblock
\APACrefYearMonthDay{2022}{11}{}.
\newblock
{\BBOQ}\APACrefatitle {{QGreenland: Enabling Science through GIS}} {{QGreenland: Enabling Science through GIS}}.{\BBCQ}
\newblock
\APACjournalVolNumPages{Authorea Preprints}{}{}{}.
\newblock
\begin{APACrefDOI} \doi{10.1002/ESSOAR.10504079.1} \end{APACrefDOI}
\PrintBackRefs{\CurrentBib}

\bibitem [\protect \citeauthoryear {%
Morlighem%
\ \protect \BOthers {.}}{%
Morlighem%
\ \protect \BOthers {.}}{%
{\protect \APACyear {2022}}%
}]{%
Morlighem2022IceBridge5}
\APACinsertmetastar {%
Morlighem2022IceBridge5}%
\begin{APACrefauthors}%
Morlighem, M.%
, Williams, C.%
, Rignot, E.%
, An, L.%
, Arndt, J.%
, Bamber, J.%
\BDBL {}Zinglersen, K.%
\end{APACrefauthors}%
\unskip\
\newblock
\APACrefYearMonthDay{2022}{}{}.
\newblock
\APACrefbtitle {{IceBridge BedMachine Greenland, Version 5}.} {{IceBridge BedMachine Greenland, Version 5}.}
\newblock
\APACaddressPublisher{Colorado, USA}{NASA National Snow and Ice Data Center Distributed Active Archive Center.}
\PrintBackRefs{\CurrentBib}

\bibitem [\protect \citeauthoryear {%
Morlighem%
\ \protect \BOthers {.}}{%
Morlighem%
\ \protect \BOthers {.}}{%
{\protect \APACyear {2017}}%
}]{%
Morlighem2017BedMachineConservation}
\APACinsertmetastar {%
Morlighem2017BedMachineConservation}%
\begin{APACrefauthors}%
Morlighem, M.%
, Williams, C\BPBI N.%
, Rignot, E.%
, An, L.%
, Arndt, J\BPBI E.%
, Bamber, J\BPBI L.%
\BDBL {}{others}%
\end{APACrefauthors}%
\unskip\
\newblock
\APACrefYearMonthDay{2017}{11}{}.
\newblock
{\BBOQ}\APACrefatitle {{BedMachine v3: Complete Bed Topography and Ocean Bathymetry Mapping of Greenland From Multibeam Echo Sounding Combined With Mass Conservation}} {{BedMachine v3: Complete Bed Topography and Ocean Bathymetry Mapping of Greenland From Multibeam Echo Sounding Combined With Mass Conservation}}.{\BBCQ}
\newblock
\APACjournalVolNumPages{Geophysical Research Letters}{44}{21}{11051--11061}.
\newblock
\begin{APACrefDOI} \doi{10.1002/2017GL074954} \end{APACrefDOI}
\PrintBackRefs{\CurrentBib}

\bibitem [\protect \citeauthoryear {%
No{\"{e}}l%
\ \protect \BOthers {.}}{%
No{\"{e}}l%
\ \protect \BOthers {.}}{%
{\protect \APACyear {2018}}%
}]{%
Noel2018Modelling19582016}
\APACinsertmetastar {%
Noel2018Modelling19582016}%
\begin{APACrefauthors}%
No{\"{e}}l, B.%
, Van De~Berg, W\BPBI J.%
, Melchior Van~Wessem, J.%
, Van~Meijgaard, E.%
, Van~As, D.%
, Lenaerts, J\BPBI T\BPBI M.%
\BDBL {}Van Den~Broeke, M\BPBI R.%
\end{APACrefauthors}%
\unskip\
\newblock
\APACrefYearMonthDay{2018}{}{}.
\newblock
{\BBOQ}\APACrefatitle {{Modelling the climate and surface mass balance of polar ice sheets using RACMO2 – Part 1: Greenland (1958–2016)}} {{Modelling the climate and surface mass balance of polar ice sheets using RACMO2 – Part 1: Greenland (1958–2016)}}.{\BBCQ}
\newblock
\APACjournalVolNumPages{The Cryosphere}{12}{}{811--831}.
\newblock
\begin{APACrefURL} \url{https://doi.org/10.5194/tc-12-811-2018} \end{APACrefURL}
\newblock
\begin{APACrefDOI} \doi{10.5194/tc-12-811-2018} \end{APACrefDOI}
\PrintBackRefs{\CurrentBib}

\bibitem [\protect \citeauthoryear {%
Nye%
}{%
Nye%
}{%
{\protect \APACyear {1952}}%
}]{%
Nye1952TheFlow}
\APACinsertmetastar {%
Nye1952TheFlow}%
\begin{APACrefauthors}%
Nye, J\BPBI F.%
\end{APACrefauthors}%
\unskip\
\newblock
\APACrefYearMonthDay{1952}{}{}.
\newblock
{\BBOQ}\APACrefatitle {{The Mechanics of Glacier Flow}} {{The Mechanics of Glacier Flow}}.{\BBCQ}
\newblock
\APACjournalVolNumPages{Journal of Glaciology}{2}{12}{82--93}.
\newblock
\begin{APACrefDOI} \doi{10.3189/S0022143000033967} \end{APACrefDOI}
\PrintBackRefs{\CurrentBib}

\bibitem [\protect \citeauthoryear {%
Otosaka%
\ \protect \BOthers {.}}{%
Otosaka%
\ \protect \BOthers {.}}{%
{\protect \APACyear {2023}}%
}]{%
Otosaka2023Mass2020}
\APACinsertmetastar {%
Otosaka2023Mass2020}%
\begin{APACrefauthors}%
Otosaka, I\BPBI N.%
, Shepherd, A.%
, Ivins, E\BPBI R.%
, Schlegel, N\BPBI J.%
, Amory, C.%
, Van Den~Broeke, M\BPBI R.%
\BDBL {}Wouters, B.%
\end{APACrefauthors}%
\unskip\
\newblock
\APACrefYearMonthDay{2023}{4}{}.
\newblock
{\BBOQ}\APACrefatitle {{Mass balance of the Greenland and Antarctic ice sheets from 1992 to 2020}} {{Mass balance of the Greenland and Antarctic ice sheets from 1992 to 2020}}.{\BBCQ}
\newblock
\APACjournalVolNumPages{Earth System Science Data}{15}{4}{1597--1616}.
\newblock
\begin{APACrefDOI} \doi{10.5194/ESSD-15-1597-2023} \end{APACrefDOI}
\PrintBackRefs{\CurrentBib}

\bibitem [\protect \citeauthoryear {%
Ozoe%
\ \BBA {} Churchill%
}{%
Ozoe%
\ \BBA {} Churchill%
}{%
{\protect \APACyear {1972}}%
}]{%
Ozoe1972HydrodynamicSolution}
\APACinsertmetastar {%
Ozoe1972HydrodynamicSolution}%
\begin{APACrefauthors}%
Ozoe, H.%
\BCBT {}\ \BBA {} Churchill, S\BPBI W.%
\end{APACrefauthors}%
\unskip\
\newblock
\APACrefYearMonthDay{1972}{}{}.
\newblock
{\BBOQ}\APACrefatitle {{Hydrodynamic stability and natural convection in Ostwald-de Waele and Ellis fluids: The development of a numerical solution}} {{Hydrodynamic stability and natural convection in Ostwald-de Waele and Ellis fluids: The development of a numerical solution}}.{\BBCQ}
\newblock
\APACjournalVolNumPages{AIChE}{18}{6}{1196--1207}.
\newblock
\begin{APACrefDOI} \doi{10.1002/AIC.690180617} \end{APACrefDOI}
\PrintBackRefs{\CurrentBib}

\bibitem [\protect \citeauthoryear {%
Panton%
\ \BBA {} Karlsson%
}{%
Panton%
\ \BBA {} Karlsson%
}{%
{\protect \APACyear {2015}}%
}]{%
Panton2015AutomatedSheet}
\APACinsertmetastar {%
Panton2015AutomatedSheet}%
\begin{APACrefauthors}%
Panton, C.%
\BCBT {}\ \BBA {} Karlsson, N\BPBI B.%
\end{APACrefauthors}%
\unskip\
\newblock
\APACrefYearMonthDay{2015}{12}{}.
\newblock
{\BBOQ}\APACrefatitle {{Automated mapping of near bed radio-echo layer disruptions in the Greenland Ice Sheet}} {{Automated mapping of near bed radio-echo layer disruptions in the Greenland Ice Sheet}}.{\BBCQ}
\newblock
\APACjournalVolNumPages{Earth and Planetary Science Letters}{432}{}{323--331}.
\newblock
\begin{APACrefDOI} \doi{10.1016/J.EPSL.2015.10.024} \end{APACrefDOI}
\PrintBackRefs{\CurrentBib}

\bibitem [\protect \citeauthoryear {%
Parmentier%
}{%
Parmentier%
}{%
{\protect \APACyear {1978}}%
}]{%
Parmentier1978AFluids}
\APACinsertmetastar {%
Parmentier1978AFluids}%
\begin{APACrefauthors}%
Parmentier, E\BPBI M.%
\end{APACrefauthors}%
\unskip\
\newblock
\APACrefYearMonthDay{1978}{}{}.
\newblock
{\BBOQ}\APACrefatitle {{A study of thermal convection in non-Newtonian fluids}} {{A study of thermal convection in non-Newtonian fluids}}.{\BBCQ}
\newblock
\APACjournalVolNumPages{Journal of Fluid Mechanics}{84}{1}{1--11}.
\newblock
\begin{APACrefDOI} \doi{10.1017/S0022112078000014} \end{APACrefDOI}
\PrintBackRefs{\CurrentBib}

\bibitem [\protect \citeauthoryear {%
Paterson%
}{%
Paterson%
}{%
{\protect \APACyear {1991}}%
}]{%
Paterson1991Whysoft}
\APACinsertmetastar {%
Paterson1991Whysoft}%
\begin{APACrefauthors}%
Paterson, W\BPBI S.%
\end{APACrefauthors}%
\unskip\
\newblock
\APACrefYearMonthDay{1991}{11}{}.
\newblock
{\BBOQ}\APACrefatitle {{Why ice-age ice is sometimes “soft”}} {{Why ice-age ice is sometimes “soft”}}.{\BBCQ}
\newblock
\APACjournalVolNumPages{Cold Regions Science and Technology}{20}{1}{75--98}.
\newblock
\begin{APACrefDOI} \doi{10.1016/0165-232X(91)90058-O} \end{APACrefDOI}
\PrintBackRefs{\CurrentBib}

\bibitem [\protect \citeauthoryear {%
Ranganathan%
\ \BBA {} Minchew%
}{%
Ranganathan%
\ \BBA {} Minchew%
}{%
{\protect \APACyear {2024}}%
}]{%
Ranganathan2024ASheets}
\APACinsertmetastar {%
Ranganathan2024ASheets}%
\begin{APACrefauthors}%
Ranganathan, M.%
\BCBT {}\ \BBA {} Minchew, B.%
\end{APACrefauthors}%
\unskip\
\newblock
\APACrefYearMonthDay{2024}{6}{}.
\newblock
{\BBOQ}\APACrefatitle {{A modified viscous flow law for natural glacier ice: Scaling from laboratories to ice sheets}} {{A modified viscous flow law for natural glacier ice: Scaling from laboratories to ice sheets}}.{\BBCQ}
\newblock
\APACjournalVolNumPages{Proceedings of the National Academy of Sciences of the United States of America}{121}{23}{e2309788121}.
\newblock
\begin{APACrefDOI} \doi{https://doi.org/10.1073/pnas.2309788121} \end{APACrefDOI}
\PrintBackRefs{\CurrentBib}

\bibitem [\protect \citeauthoryear {%
Rasmussen%
\ \protect \BOthers {.}}{%
Rasmussen%
\ \protect \BOthers {.}}{%
{\protect \APACyear {2013}}%
}]{%
Rasmussen2013ACore}
\APACinsertmetastar {%
Rasmussen2013ACore}%
\begin{APACrefauthors}%
Rasmussen, S\BPBI O.%
, Abbott, P\BPBI M.%
, Blunier, T.%
, Bourne, A\BPBI J.%
, Brook, E.%
, Buchardt, S\BPBI L.%
\BDBL {}Winstrup, M.%
\end{APACrefauthors}%
\unskip\
\newblock
\APACrefYearMonthDay{2013}{12}{}.
\newblock
{\BBOQ}\APACrefatitle {{A first chronology for the north greenland eemian ice drilling (NEEM) ice core}} {{A first chronology for the north greenland eemian ice drilling (NEEM) ice core}}.{\BBCQ}
\newblock
\APACjournalVolNumPages{Climate of the Past}{9}{6}{2713--2730}.
\newblock
\begin{APACrefDOI} \doi{10.5194/CP-9-2713-2013} \end{APACrefDOI}
\PrintBackRefs{\CurrentBib}

\bibitem [\protect \citeauthoryear {%
Rayleigh%
}{%
Rayleigh%
}{%
{\protect \APACyear {1916}}%
}]{%
Rayleigh1916LIX.Side}
\APACinsertmetastar {%
Rayleigh1916LIX.Side}%
\begin{APACrefauthors}%
Rayleigh, O.%
\end{APACrefauthors}%
\unskip\
\newblock
\APACrefYearMonthDay{1916}{}{}.
\newblock
{\BBOQ}\APACrefatitle {{LIX. On convection currents in a horizontal layer of fluid, when the higher temperature is on the under side}} {{LIX. On convection currents in a horizontal layer of fluid, when the higher temperature is on the under side}}.{\BBCQ}
\newblock
\APACjournalVolNumPages{The London, Edinburgh, and Dublin Philosophical Magazine and Journal of Science}{}{}{}.
\newblock
\begin{APACrefDOI} \doi{10.1080/14786441608635602} \end{APACrefDOI}
\PrintBackRefs{\CurrentBib}

\bibitem [\protect \citeauthoryear {%
Rieckh%
, Born%
, Robinson%
, Law%
\BCBL {}\ \BBA {} G{\"{u}}lle%
}{%
Rieckh%
\ \protect \BOthers {.}}{%
{\protect \APACyear {2024}}%
}]{%
Rieckh2024DesignTracing}
\APACinsertmetastar {%
Rieckh2024DesignTracing}%
\begin{APACrefauthors}%
Rieckh, T.%
, Born, A.%
, Robinson, A.%
, Law, R.%
\BCBL {}\ \BBA {} G{\"{u}}lle, G.%
\end{APACrefauthors}%
\unskip\
\newblock
\APACrefYearMonthDay{2024}{9}{}.
\newblock
{\BBOQ}\APACrefatitle {{Design and performance of ELSA v2.0: an isochronal model for ice-sheet layer tracing}} {{Design and performance of ELSA v2.0: an isochronal model for ice-sheet layer tracing}}.{\BBCQ}
\newblock
\APACjournalVolNumPages{Geoscientific Model Development}{17}{18}{6987--7000}.
\newblock
\begin{APACrefDOI} \doi{10.5194/GMD-17-6987-2024} \end{APACrefDOI}
\PrintBackRefs{\CurrentBib}

\bibitem [\protect \citeauthoryear {%
Sanderson%
\ \protect \BOthers {.}}{%
Sanderson%
\ \protect \BOthers {.}}{%
{\protect \APACyear {2023}}%
}]{%
Sanderson2023EnglacialAntarctica}
\APACinsertmetastar {%
Sanderson2023EnglacialAntarctica}%
\begin{APACrefauthors}%
Sanderson, R\BPBI J.%
, Winter, K.%
, Callard, S\BPBI L.%
, Napoleoni, F.%
, Ross, N.%
, Jordan, T\BPBI A.%
\BCBL {}\ \BBA {} Bingham, R\BPBI G.%
\end{APACrefauthors}%
\unskip\
\newblock
\APACrefYearMonthDay{2023}{11}{}.
\newblock
{\BBOQ}\APACrefatitle {{Englacial architecture of Lambert Glacier, East Antarctica}} {{Englacial architecture of Lambert Glacier, East Antarctica}}.{\BBCQ}
\newblock
\APACjournalVolNumPages{Cryosphere}{17}{11}{4853--4871}.
\newblock
\begin{APACrefDOI} \doi{10.5194/TC-17-4853-2023} \end{APACrefDOI}
\PrintBackRefs{\CurrentBib}

\bibitem [\protect \citeauthoryear {%
Solomatov%
}{%
Solomatov%
}{%
{\protect \APACyear {1995}}%
}]{%
Solomatov1995ScalingConvection}
\APACinsertmetastar {%
Solomatov1995ScalingConvection}%
\begin{APACrefauthors}%
Solomatov, V\BPBI S.%
\end{APACrefauthors}%
\unskip\
\newblock
\APACrefYearMonthDay{1995}{2}{}.
\newblock
{\BBOQ}\APACrefatitle {{Scaling of temperature‐ and stress‐dependent viscosity convection}} {{Scaling of temperature‐ and stress‐dependent viscosity convection}}.{\BBCQ}
\newblock
\APACjournalVolNumPages{Physics of Fluids}{7}{2}{266--274}.
\newblock
\begin{APACrefURL} \url{/aip/pof/article/7/2/266/258803/Scaling-of-temperature-and-stress-dependent} \end{APACrefURL}
\newblock
\begin{APACrefDOI} \doi{10.1063/1.868624} \end{APACrefDOI}
\PrintBackRefs{\CurrentBib}

\bibitem [\protect \citeauthoryear {%
Talalay%
\ \BBA {} Hooke%
}{%
Talalay%
\ \BBA {} Hooke%
}{%
{\protect \APACyear {2007}}%
}]{%
Talalay2007ClosureDiscussion}
\APACinsertmetastar {%
Talalay2007ClosureDiscussion}%
\begin{APACrefauthors}%
Talalay, P\BPBI G.%
\BCBT {}\ \BBA {} Hooke, R\BPBI L\BPBI B.%
\end{APACrefauthors}%
\unskip\
\newblock
\APACrefYearMonthDay{2007}{}{}.
\newblock
{\BBOQ}\APACrefatitle {{Closure of deep boreholes in ice sheets: a discussion}} {{Closure of deep boreholes in ice sheets: a discussion}}.{\BBCQ}
\newblock
\APACjournalVolNumPages{Annals of Glaciology}{47}{}{125--133}.
\newblock
\begin{APACrefDOI} \doi{10.3189/172756407786857794} \end{APACrefDOI}
\PrintBackRefs{\CurrentBib}

\bibitem [\protect \citeauthoryear {%
Theofilopoulos%
\ \BBA {} Born%
}{%
Theofilopoulos%
\ \BBA {} Born%
}{%
{\protect \APACyear {2023}}%
}]{%
Theofilopoulos2023SensitivityDynamics}
\APACinsertmetastar {%
Theofilopoulos2023SensitivityDynamics}%
\begin{APACrefauthors}%
Theofilopoulos, A.%
\BCBT {}\ \BBA {} Born, A.%
\end{APACrefauthors}%
\unskip\
\newblock
\APACrefYearMonthDay{2023}{4}{}.
\newblock
{\BBOQ}\APACrefatitle {{Sensitivity of isochrones to surface mass balance and dynamics}} {{Sensitivity of isochrones to surface mass balance and dynamics}}.{\BBCQ}
\newblock
\APACjournalVolNumPages{Journal of Glaciology}{69}{274}{311--323}.
\newblock
\begin{APACrefDOI} \doi{10.1017/JOG.2022.62} \end{APACrefDOI}
\PrintBackRefs{\CurrentBib}

\bibitem [\protect \citeauthoryear {%
Whillans%
}{%
Whillans%
}{%
{\protect \APACyear {1977}}%
}]{%
Whillans1977TheAntarctica}
\APACinsertmetastar {%
Whillans1977TheAntarctica}%
\begin{APACrefauthors}%
Whillans, I\BPBI M.%
\end{APACrefauthors}%
\unskip\
\newblock
\APACrefYearMonthDay{1977}{}{}.
\newblock
{\BBOQ}\APACrefatitle {{The Equation of Continuity and its Application to the Ice Sheet Near “byrd” Station, Antarctica}} {{The Equation of Continuity and its Application to the Ice Sheet Near “byrd” Station, Antarctica}}.{\BBCQ}
\newblock
\APACjournalVolNumPages{Journal of Glaciology}{18}{80}{359--371}.
\newblock
\begin{APACrefDOI} \doi{10.3189/S0022143000021055} \end{APACrefDOI}
\PrintBackRefs{\CurrentBib}

\bibitem [\protect \citeauthoryear {%
Wolovick%
, Creyts%
, Buck%
\BCBL {}\ \BBA {} Bell%
}{%
Wolovick%
\ \protect \BOthers {.}}{%
{\protect \APACyear {2014}}%
}]{%
Wolovick2014TravelingSheets}
\APACinsertmetastar {%
Wolovick2014TravelingSheets}%
\begin{APACrefauthors}%
Wolovick, M\BPBI J.%
, Creyts, T\BPBI T.%
, Buck, W\BPBI R.%
\BCBL {}\ \BBA {} Bell, R\BPBI E.%
\end{APACrefauthors}%
\unskip\
\newblock
\APACrefYearMonthDay{2014}{12}{}.
\newblock
{\BBOQ}\APACrefatitle {{Traveling slippery patches produce thickness-scale folds in ice sheets}} {{Traveling slippery patches produce thickness-scale folds in ice sheets}}.{\BBCQ}
\newblock
\APACjournalVolNumPages{Geophysical Research Letters}{41}{24}{8895--8901}.
\newblock
\begin{APACrefDOI} \doi{10.1002/2014GL062248} \end{APACrefDOI}
\PrintBackRefs{\CurrentBib}

\bibitem [\protect \citeauthoryear {%
Zeitz%
, Levermann%
\BCBL {}\ \BBA {} Winkelmann%
}{%
Zeitz%
\ \protect \BOthers {.}}{%
{\protect \APACyear {2020}}%
}]{%
Zeitz2020SensitivityGeometry}
\APACinsertmetastar {%
Zeitz2020SensitivityGeometry}%
\begin{APACrefauthors}%
Zeitz, M.%
, Levermann, A.%
\BCBL {}\ \BBA {} Winkelmann, R.%
\end{APACrefauthors}%
\unskip\
\newblock
\APACrefYearMonthDay{2020}{10}{}.
\newblock
{\BBOQ}\APACrefatitle {{Sensitivity of ice loss to uncertainty in flow law parameters in an idealized one-dimensional geometry}} {{Sensitivity of ice loss to uncertainty in flow law parameters in an idealized one-dimensional geometry}}.{\BBCQ}
\newblock
\APACjournalVolNumPages{Cryosphere}{14}{10}{3537--3550}.
\newblock
\begin{APACrefDOI} \doi{10.5194/TC-14-3537-2020} \end{APACrefDOI}
\PrintBackRefs{\CurrentBib}

\bibitem [\protect \citeauthoryear {%
T.~Zhang%
\ \protect \BOthers {.}}{%
T.~Zhang%
\ \protect \BOthers {.}}{%
{\protect \APACyear {2024}}%
}]{%
Zhang2024EvaluatingSheet}
\APACinsertmetastar {%
Zhang2024EvaluatingSheet}%
\begin{APACrefauthors}%
Zhang, T.%
, Colgan, W.%
, Wansing, A.%
, L{\o}kkegaard, A.%
, Leguy, G.%
, Lipscomb, W\BPBI H.%
\BCBL {}\ \BBA {} Xiao, C.%
\end{APACrefauthors}%
\unskip\
\newblock
\APACrefYearMonthDay{2024}{1}{}.
\newblock
{\BBOQ}\APACrefatitle {{Evaluating different geothermal heat-flow maps as basal boundary conditions during spin-up of the Greenland ice sheet}} {{Evaluating different geothermal heat-flow maps as basal boundary conditions during spin-up of the Greenland ice sheet}}.{\BBCQ}
\newblock
\APACjournalVolNumPages{Cryosphere}{18}{1}{387--402}.
\newblock
\begin{APACrefDOI} \doi{10.5194/TC-18-387-2024} \end{APACrefDOI}
\PrintBackRefs{\CurrentBib}

\bibitem [\protect \citeauthoryear {%
Y.~Zhang%
\ \protect \BOthers {.}}{%
Y.~Zhang%
\ \protect \BOthers {.}}{%
{\protect \APACyear {2024}}%
}]{%
Zhang2024FormationSheet}
\APACinsertmetastar {%
Zhang2024FormationSheet}%
\begin{APACrefauthors}%
Zhang, Y.%
, Sachau, T.%
, Franke, S.%
, Yang, H.%
, Li, D.%
, Weikusat, I.%
\BCBL {}\ \BBA {} Bons, P\BPBI D.%
\end{APACrefauthors}%
\unskip\
\newblock
\APACrefYearMonthDay{2024}{8}{}.
\newblock
{\BBOQ}\APACrefatitle {{Formation Mechanisms of Large-Scale Folding in Greenland's Ice Sheet}} {{Formation Mechanisms of Large-Scale Folding in Greenland's Ice Sheet}}.{\BBCQ}
\newblock
\APACjournalVolNumPages{Geophysical Research Letters}{51}{16}{e2024GL109492}.
\newblock
\begin{APACrefDOI} \doi{10.1029/2024GL109492} \end{APACrefDOI}
\PrintBackRefs{\CurrentBib}

\end{thebibliography}

\end{document}